\documentclass[aps,prb,twocolumn,nopacs,amssymb]{revtex4}
\usepackage[dvips]{graphicx}
\usepackage{dcolumn}
\usepackage{bm}
\def\gs{\gtrsim}
\def\ls{\lesssim}
\def\be{\begin{equation}}
\def\en{\end{equation}}    
\def\gs{\gtrsim}
\def\ls{\lesssim}
\newcommand{\bi}[1]{\mbox{\boldmath$#1$}}

\def\p{\partial}
\def\bea{\begin{eqnarray}}
\def\ena{\end{eqnarray}}
\newcommand{\ppp}[3]{{\bigg(}\frac{\partial {#1}}{\partial {#2}}{\bigg )}_{#3}}
\renewcommand{\theequation}{\arabic{section}.\arabic{equation}}

\begin{document}
\draft
\bibliographystyle{prsty}
\title{Casimir amplitudes  and 
capillary condensation of   near-critical  fluids  
between  parallel plates: 
 Renormalized local functional  theory  }
\author{Ryuichi Okamoto  and Akira Onuki}
\address{Department of Physics, Kyoto University, Kyoto 606-8502}
\date{\today}

\begin{abstract} 
We investigate the critical behavior   of 
 a near-critical fluid  
 confined between two parallel plates 
in contact with a reservoir 
by  calculating  the order parameter  profile and  
the Casimir amplitudes (for the force density 
and   for the grand potential). 
Our results are applicable 
to one-component fluids and binary mixtures. 
We assume that the walls 
  absorb one of the fluid components selectively for 
binary mixtures.    We propose  a renormalized local 
 functional theory accounting  for the fluctuation effects. 
 Analysis is performed in the plane of the 
 temperature $T$ and 
 the order parameter in the reservoir $\psi_\infty$.    
Our theory is universal if the physical quantities are scaled 
appropriately.  If the component favored by the walls 
is slightly poor   in the reservoir, 
there appears  a line of first-order phase transition 
of capillary condensation outside the bulk coexistence  curve. 
The excess adsorption changes discontinuously  
between   condensed and  noncondensed  states at  the transition. 
With increasing $T$, the transition line ends at a capillary 
critical point $T=T_c^{\rm ca}$ slightly lower than 
the bulk critical temperature $T_c$. 
The Casimir amplitudes  
are  larger than 
their critical-point   values  
by 10-100 times at off-critical compositions near the 
capillary condensation line.  
\end{abstract}


\maketitle


\section{Introduction}

 Understanding the  phase behavior of fluids  
confined  in narrow regions 
is crucial  in the physics of fluids in porous 
media  and for surface force experiments \cite{Evansreview,Gelb}. 
It  strongly depends on   the molecular 
interactions. In  binary mixtures, 
one of the solvent components 
is preferentially attracted 
to the wall 
\cite{Cahn,PG}. 
 In liquid water in contact with  a hydrophobic  surface, 
water molecules tend to be separated from 
the surface due to the hydrogen bonding 
 among the water molecules,  
resulting  in the 
formation of  a  gaseous layer on the surface 
 \cite{Chandler}.   
In aqueous mixtures with salt, 
  prewetting behavior is much intensified by 
  the composition-dependent  surface ionization 
and the  preferential   solvation   of ions 
\cite{Onuki-Kitamura,Current,Okamoto}.  In these examples,  
heterogeneities in  the density 
or the composition are induced near the wall, 
often resulting in wetting or drying transition 
on  the wall. The confinement effects 
can be dramatic  
when the spatial scale of the fluid heterogeneities 
is on the order of the  wall separation.

Narrow regions may be  
filled with the phase favored by  the  confining 
walls or may hold some fraction of 
the disfavored  phase 
\cite{Evansreview,Gelb,Nakanishi-s,Nakanishi-mean,Evans-Marconi,Evans-Anna}, 
depending  on   the  temperature $T$, 
and the reservoir chemical potential 
difference 
 $\mu_\infty$ \cite{chemical}
 (or the reservoir order parameter $\psi_\infty$) 
for each given wall separation $D$   
between these two states, 
there is  a first-order 
phase transition line 
slightly outside the bulk coexistence curve in the 
$T$-$\mu_\infty$ plane.  In this paper, we call 
it  a capillary condensation line, 
though this name is usually used  
for gas-liquid phase separation in narrow regions. 
For binary mixtures, 
there is a discontinuity in the preferential adsorption 
and the osmotic pressure across this line. 
Macio\l ek {\it et al.} \cite{Evans-Anna} 
numerically found a capillary condensation line in 
two-dimensional Ising films.  
Samin and Tsori \cite{Tsori} 
found  a first-order transition 
 in  binary mixtures  with ions  between parallel 
 plates  including  the selective solvation 
 in the mean-field theory,  
  where the osmotic  pressure is discontinuous. 
Our aim in this paper is  to investigate 
this  capillary condensation transition 
in near-critical fluids 
 between parallel plates,   
taking into account the renormalization effect of 
the critical fluctuations. 
For simplicity, no ions will be assumed.

In  near-critical  fluids,   
 the wall-induced heterogeneities extend 
 over mesoscopic length scales,   
 leading to an intriguing interplay between the 
 finite-size effect and the molecular interactions  
 \cite{Fisher,Fisher-Yang}. 
Note that various    surface phase transitions  
 are known near the bulk criticality 
 depending  on the type of surface 
 \cite{Nakanishi,Binderreview,Dietrichreview}. 
According to the prediction by Fisher and de Gennes,  
the  free energy  
of a film with thickness $D$ and area $S$ 
at the bulk criticality 
consists of a bulk part proportional to the volume $SD $,  
 wall contributions proportional to the area $S$, 
and the interaction part, 
 \cite{Ha,In,K-D,Krechreview,Krech,JSP,Gamb,Monte,Upton,Upton1},
 \be 
 \Delta F= - S k_BT_c  D^{-d+1}\Delta,
\en   
where  $d$ is the space 
dimensionality, and $\Delta$ is a universal number 
only depending on the type of the boundary conditions. 
In the literature, $\Delta$ 
has been called the  the  Casimir amplitude \cite{Casimir}. 
Positivity (negativity) 
of $\Delta$ implies attractive (repulsive) interaction 
between the plates. The force density between the plates 
is then written as 
\be 
\frac{\p }{\p D}\Delta F= S{k_BT_c}D^{-d} 
{\cal A}, 
\en    
where $\cal A$ is another  amplitude. 
The amplitudes $\Delta$ and 
 $\cal A$  depend on the  
reduced temperature $\tau= T/T_c-1$ and 
the reservoir chemical potential difference $\mu_\infty$.  
However,  $\Delta$  
and $\cal A$ have been measured  as functions of $\tau$  
 along the  critical path $\mu_\infty=0$ 
 \cite{Law,Lawreview,Rafai}.
In this paper, we   calculate them 
in  the $\tau$-$\mu_\infty$  plane 
 to find   their dramatic enhancement 
 near  the capillary condensation line, 
 where the reservoir composition 
 is poor in the component favored by the walls. 
 In    accord with our  result,    $\cal A$ increased 
 dramatically  around a capillary critical line     
  under  a magnetic field in 
 two-dimensional Ising films        
 \cite{Evans-Anna}.

  In  binary mixtures, 
the adsorption-induced 
composition disturbances 
   produce  an attractive interaction  
 between solid objects 
 \cite{Evans-Hop,Okamoto}.
 Indeed, reversible aggregation of colloid particles 
 have been observed 
 close to the critical point at off-critical 
 compositions \cite{Beysens,Maher,Kaler,Bonn,Guo}. 
 On approaching 
the solvent criticality, 
 this  interaction  becomes  long-ranged and 
  universal in the  strong adsorption limit 
\cite{Lawreview,Fisher-Yang}. 
 Recently, the near-critical 
colloid-wall interaction has been  
measured directly  at the critical composition 
\cite{Nature2008,salt-soft}.   
However, the interaction should be  much larger at 
off-critical compositions.  We mention 
a  microscopic theory  by 
Hopkins {\it et al.} \cite{Evans-Hop}, 
who   found enhancement of 
 the  colloid interaction  in one-phase environments 
 poor  in  the component favorted by the 
 colloid surface. 
    Furthermore,   ions can strongly affect 
the colloid interaction  in an aqueous  mixture   \cite{Okamoto},  
since  the ion  distributions become    
highly heterogeneous near the charged 
surface inducing  formation of a wetting layer 
on the surface.

The organization of this paper is as follows. 
In Sec.II,  we will present   a  local functional theory 
 of  a binary mixture  
at the critical temperature $\tau=T/T_c-1=0$. 
The Casimir amplitudes can easily be calculated from this model. 
In Sec.III, we will extend our  theory 
in Sec.II to the case of 
 nonvanishing reduced temperature $\tau$ 
  accounting for the renormalization effect.
We can  then  calculate  the Casimir amplitudes 
 for nonvanishing $\tau$ and $\mu_\infty$ 
 and predict the capillary condensation 
transition.

\section{Critical behavior at $T=T_c$}
\setcounter{equation}{0}

In this section, 
we treat   a near-critical binary mixture 
 at the critical temperature $T=T_c$ at a fixed pressure 
 without ions. 
 Using  the  free energy proposed by 
 Fisher and Au-Yang \cite{Fisher-Yang} at $T=T_c$, 
 we  calculate the composition profiles 
 and the Casimir amplitudes  
 in the semi-infinite and film geometries.  
 We mention similar calculations   by 
 Borjan and Upton 
\cite{Upton,Upton1,Upton-ad} along the critical path $\mu_\infty=0$.  
The method in this section 
will be generalized  
to  the case of  nonvanishing $\mu_\infty$ 
and $T-T_c$ in the next section.

\subsection{Model of Fisher and Au-Yang }

The order parameter is the deviation $\psi=c-c_c$ 
of the composition $c$ from its critical value $c_c$.  
Supposing spatial 
 variations of $\psi$ on scales much longer than 
 the molecular diameter,  we  use the following 
form for the  singular free energy 
density including the gradient 
contribution  \cite{Fisher-Yang}, 
\be 
f_{\rm loc} =   f(\psi)+  
\frac{1}{2}C(\psi) 
|\nabla \psi|^2, 
\en    
The total singular free energy is given by 
\be 
F=\int d{\bi r}[ f_{\rm loc}-\mu_\infty\psi]
+ \int dS f_s(\psi),
\en 
where  $\mu_\infty$ is a given 
 chemical potential difference  
in  the  reservoir   \cite{chemical} 
(magnetic field $h$ for Ising spin systems).  
The second term is the surface integral of 
 the surface free energy  $f_s(\psi)$ 
arising from the short-range interaction between 
the fluid and the wall \cite{Cahn} (with $dS$ being the surface 
element).   At $T=T_c$,   $f(\psi) $ and $C(\psi)$ are proportional to 
 fractional powers of $|\psi|$ as  
\bea 
f(\psi) &=& {B_0} |\psi|^{1+\delta}, \\
C(\psi) &=&  C_0   |\psi|^{-\eta\nu/\beta}, 
\ena 
where  $B_0$ and $C_0$ are 
positive constants.   
The   $\delta$, 
$\eta$, $\nu$, and  $\beta$ 
 are the usual critical exponents 
for Ising-like systems. 
For the spatial dimensionality 
$d$ in the range $2\le d\le 4$, 
they satisfy the exponent relations\cite{Onukibook}, 
\be 
\delta= \frac{d+2-\eta}{d-2+\eta} 
= \frac{d\nu}{\beta}-1, \quad 
\frac{\beta}{\nu}= \frac{d-2+\eta}{2}.
\en  
The exponent $\eta$ is very small in the range $0.03-0.04$ 
for $d=3$.  The other exponenta we will use are $\gamma=
(2-\eta)\nu= (\delta-1)\beta$ and $\alpha=2-\nu d$. 
In our numerical analysis to follow, 
we  set 
$\alpha=2-\nu d= 0.11$, $\beta=0.325$,  
$\delta=4.815$, $\nu=0.630$, and $\eta=0.0317$. 
They  are three-dimensional values 
 satisfying  Eq.(2.5).

We define the local correlation 
length $\xi_{\rm loc}(\psi)$   by 
\be 
\xi_{\rm loc}(\psi)= (2C_0/B_0)^{1/2} |\psi|^{-\nu/\beta}. 
\en 
The two terms in Eq.(2.1) 
are of the same order if we set $|\nabla \psi|^2\sim 
\psi^2/\xi_{\rm loc}^2$. 
The combination 
$(2C_0/B_0)^{1/2}$ is  a microscopic length 
if $\psi$ denotes  the composition deviation. 
In terms of   $\xi_{\rm loc}$, we rewrite Eq.(2.1) as \cite{Fisher-Yang}
\be 
\xi_{\rm loc}^{d}
f_{\rm loc}/k_BT_c = A_c  
[ 1+ \xi_{\rm loc}^2|\nabla \psi|^2/4\psi^2]. 
\en   
Here  the coefficient $A_c$  is  expressed as 
\bea
A_c& =& B_0(2C_0/B_0)^{d/2} /k_BT_c.
 \nonumber\\
&=& (4C_0^2/B_0k_BT_c )(B_0/2C_0)^{\epsilon/2} .
\ena  
Hereafter  $\epsilon=4-d$. 
In this model free energy, the critical 
fluctuations with wave lengths shorter 
 than $\xi_{\rm loc}$ have already been 
renormalized or coarse-grained. Thus  minimizing $F$ in Eq.(2.2)  
yields  the average profile of $\psi$ 
near  the walls, where we neglect  the thermal fluctuations 
with   wavelengths shorter  than $\xi_{\rm loc}$. 
 The theory of the two-scale-factor universality 
 indicates that  $A_c$ in Eq.(2.8) 
 should be a universal number 
(independent of the mixture species)\cite{Stauffer,HAHS}.
The second line of Eq.(2.8) 
indicates that its  $\epsilon$ expansion form is 
$A_c= 18/\pi^2\epsilon+\cdots$ 
 as $\epsilon \to 0$. 
In the next section below Eq.(3.16), 
we will   estimate $A_c$ to be 1.49 at $d=3$.  

For a thin  film 
with thickness $D$, 
however,  the local forms in Eqs.(2.3) and (2.4) 
are  not  justified 
in  the spatial region where 
$\xi_{\rm loc}(\psi)\gg D$ holds, 
since the length of the critical fluctuations 
in the perpendicular direction cannot exceed $D$. 
Here we should note that  two-dimensional 
critical fluctuations varying in the lateral plane 
emerge near the critical point of capillary condensation, 
though they are neglected in this paper (see Sec.III). 
From the balance $\xi_{\rm loc}(\psi_D)= D$,   
we may introduce 
a   characteristic order parameter $\psi_D$  by  
\be 
\psi_D= (2C_0/B_0)^{\beta/2\nu }D^{-\beta/\nu }.
 \en  
Then $|\psi|/\psi_D= (D/\xi_{\rm loc})^{\beta/\nu}$. 
In the film geometry $\psi$ will be measured in units of $\psi_D$. 

\subsection{Profiles in the semi-infinite geometry }
 
We   first 
consider the semi-infinite system 
($0<z<\infty$)   to seek 
the one-dimensional profile $\psi= \psi(z)$. 
Here, $\psi(0)= \psi_0>0$ at $z=0$ 
and $\psi(z)\to \psi_\infty$ as $z\to \infty$. 
We treat $\psi_0$ and $\psi_\infty$  as  given parameters  
 not explicitly imposing  the usual  boundary condition 
 $C\p \psi/\p z= f_s'$ 
 at $z=0$ \cite{Cahn},   
where $f_s'= \p f_s/\p \psi$ 
with $f_s$ being  
 the surface free energy density in Eq.(2.2) 
  (see comments below Eq.(2.17)). 
We shall see how the strong adsorption limit 
 $\psi_0 \to \infty$ is attained 
near the criticality.    

The chemical potential far from the wall 
is written as   
\be 
\mu_\infty= \mu(\psi_\infty)\nonumber\\ 
=  B_0(1+\delta)|\psi_\infty|^{\delta-1} 
 \psi_\infty, 
\en 
where the  second line 
follows from Eq.(2.3). 
In  equilibrium, we minimize $F$ to obtain
\be 
\mu(\psi) -C(\psi)\psi'' - \frac{1}{2}C'(\psi) 
|\psi'|^2= \mu_\infty  ,
\en  
 where  $\psi'=d\psi/dz$,  $\psi''= 
 d^2\psi/dz^2$, and $C'(\psi) = dC/d\psi$.   
This equation is integrated to give \cite{Onukibook} 
\be 
C(\psi)|\psi'|^2=
2\omega_s(\psi). 
\en  
Here $\omega_s(\psi)$ 
is the excess grand potential density   
in the semi-infinite case defined by   
\be 
\omega_s(\psi) = 
f(\psi)-f_\infty - \mu_\infty (\psi-\psi_\infty).
\en  
where $f_\infty=f(\psi_\infty)$. 
We  further integrate Eq.(2.12) as  
\be 
z= \int_{\psi}^{\psi_0}d\psi \sqrt{C(\psi)/2\omega_s(\psi)},
\en 
where the upper bound is 
$\psi_0=\psi(0)$ and  
the lower bound tends to $\psi_\infty$ as $z$ increases.  

In particular, for $\psi_\infty=0$ or at the criticality 
in the bulk, we simply have $\omega_s= f= B_0|\psi|^{1+\delta}$. 
We may readily solve Eq.(2.12) as \cite{Nakanishi-s,Rudnick} 
\bea 
&&\hspace{-1cm}
\psi(z)= \psi_0/[ 1+ z/\ell_0]^{\beta/\nu}\nonumber\\
&&\hspace{-3mm}= (\beta/2\nu)^{\beta/\nu}
(2C_0/B_0)^{\beta/2\nu}( z+\ell_0)^{-\beta/\nu} ,
\ena  
where  $\ell_0$ is  the transition length related to $\psi_0$ by  
\be 
\ell_0= (\beta/\nu)(C_0/2B_0)^{1/2}\psi_0^{-\nu/\beta}. 
\en  
From Eq.(2.6) we have 
$\ell_0=(\beta/2\nu)\xi_{\rm loc}(\psi_0)$. 
From the second line of Eq.(2.15), 
$\psi(z)$ is independent of $\psi_0$ 
 and decays slowly   
 as $z^{-\beta/\nu}$  for $z\gg \ell_0$.   
On the other hand, the  local free energy density 
$f_{\rm loc}$ decays as  
\be 
f_{\rm loc}/ k_BT_c  = 2 (\beta/2\nu)^{d} A_c  (z+\ell_0)^{-d}.  
\en 
In the right hand side, 
$A_c$ is the universal number in  Eq.(2.8) and 
hence  the coefficient in front of 
$ (z+\ell_0)^{-d}$ is  universal. 
If  we  assume the linear form 
$f_s= -h_1 \psi(0)$, where $h_1$ is called the surface field. 
From the profile (2.15) $h_1$ and $\psi_0$ are related by 
$h_1= (\beta/\nu) C(\psi_0) \phi_0/\ell_0$.

On the other hand, 
in the off-critical case $\psi_\infty \neq 0$,  
$\psi(z)$  approaches  $\psi_\infty$ exponentially 
for large $z$ as 
\be 
\psi(z)- \psi_\infty
\sim |\psi_\infty| \exp({-z/\xi_\infty}). 
\en  
We introduce the correlation length 
 $\xi_\infty=(C/f'')^{1/2}$ in the bulk 
 region at $\psi=\psi_\infty$, where 
  $f''=\p^2 f/\p^2\psi$. 
From Eq.(2.3) it is calculated as 
\bea 
\xi_\infty&=& [C_0/\delta(\delta+1)B_0]^{1/2}
|\psi_\infty|^{-\nu/\beta} \nonumber\\
&=& \xi_{\rm loc}(\psi_\infty)
/\sqrt{2\delta(\delta+1)}
\ena  
The second line is written in terms of 
$\xi_{\rm loc} $ in Eq.(2.6). 
In Fig.1, we show the scaled profile 
$\psi(z)/|\psi_\infty|$ 
vs $z/\xi_\infty$  at $\psi_0=20 |\psi_\infty|$.
For small $\psi_\infty$, 
$\psi(z)$ approaches $|\psi_\infty|$ 
at $z\sim \xi_\infty$. 
For $\psi_\infty>0$, 
the changeover from the algebraic decay 
 to the exponential decay then takes place.
For   $\psi_\infty<0$, 
$\psi(z)$ further changes from positive to negative 
on the scale of  $\xi_\infty$.  
The  length  $\xi_\infty$ is 
proportional to $|\psi_\infty|^{-\nu/\beta}$, 
so it  becomes  longer with decreasing $|\psi_\infty|$.

 In the off-critical semi-infinite case, 
 the excess adsorption $\Gamma_{{\rm sem} \pm}
=\int_0^\infty dz (\psi(z)-\psi_\infty)$  
is finite, where the subscript 
$\pm$ represents the sign of $\psi_\infty$. 
Numerically we  find
\be
\Gamma_{{\rm sem} \pm} = 
{D}\psi_D\bigg [ B_{\pm} |\psi_D/\psi_\infty|^{k}-
(\psi_D/\psi_0)^{k}/{2k}\bigg],
\en   
where $k= \nu/\beta-1$,   
$B_+ =0.310$,  $B_-= 1.043$, and 
 $\psi_D$ is defined in Eq.(2.9). 
This formula includes the  
correction due to finite $\psi_0$ and, as a result, 
it holds within $0.1\%$ for $|\psi_0/\psi_\infty|>20$. 
We also  recognize that 
 $\Gamma_{{\rm sem}-}$ is three times 
larger than $\Gamma_{{\rm sem}+}$ for 
the same $|\psi_\infty|$. The 
 excess adsorption is larger  
for $\psi_\infty<0$	 
than for $\psi_\infty>0$. 
In this relation and those to follow, 
we may push  $\psi_0$  to infinity to obtain 
the asymptotic relations near the criticality. 

So far, $\psi_0$ 
 is assumed to be small, so  
   the transition length is given by $\ell_0$ in Eq.(2.16). 
For large $\psi_0$  of order unity, 
$\psi(z)$ should decay  into  small near-critical 
 values  if  $z$ exceeds a microscopic  
 distance.

\begin{figure}[htbp]
\begin{center}
\includegraphics[scale=0.62]{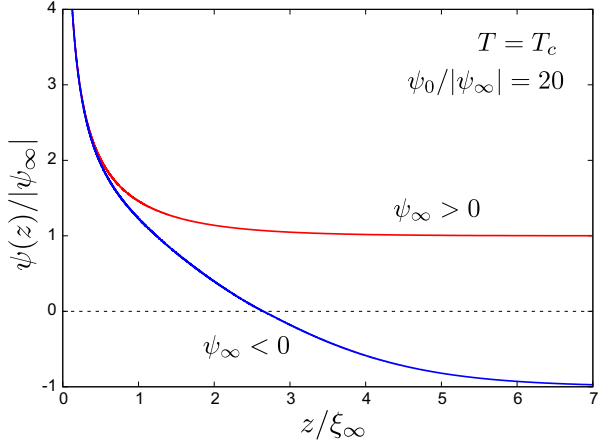}
\caption{\protect
(Color online) 
Normalized order parameter 
$\psi(z)/|\psi_\infty|$ 
vs $z/\xi_\infty$  at $T=T_c$ in the semi-infinite case 
 at $\psi_0=20 |\psi_\infty|$ 
for positive and negative small $\psi_\infty$. 
The decay is algebraic for small $z$ as 
in Eq.(2.15) and is eventually exponential as in Eq.(2.18).  
For $\psi_\infty<0$, the approach to $\psi_\infty$ is slow 
and the adsorption is large as in Eq.(2.20). 
}
\end{center}
\end{figure}

 \subsection{Profiles between parallel plates}
 
 We  assume that 
 a mixture at $T=T_c$  is  inserted between parallel 
 plates separated by $D$. The  plate area $S$ is assumed to be much 
 larger than $D^2$ such that the edge effect is negligible. 
 The fluid is   
 in contact with  a large reservoir   
 containing the same binary mixture 
 in equilibrium. 
 In the reservoir,  the  mean order parameter 
 is $\psi_\infty$ and 
 the chemical potential  
 is $\mu_\infty$ in Eq.(2.10). 
Then we  minimize  the excess grand 
potential (per unit area), 
\be 
\Omega=\int_0^D dz [  f_{\rm loc} -f_\infty  
 - \mu_\infty (\psi-\psi_\infty)]
 \en 
The fluid is in the region $0<z<D$. 
 At the walls  $z=0$ and $D$, we assume 
the symmetric boundary conditions, 
\be 
\psi(0)=\psi_0, \quad   
\psi(D)=\psi_0,   
\en 
where  $\psi_0>0$. Here  Eq.(2.11) still holds, 
leading to    
\be 
C(\psi) |d\psi/dz|^2
= 2\omega, 
\en 
where   $\omega$  is the excess grand potential 
density for a film,  
\bea 
\omega &=&  f(\psi)-f_m 
 - \mu_\infty (\psi-\psi_m) \nonumber\\   
&=&\omega_s+  f_m-f_\infty- 
 \mu_\infty (\psi_m-\psi_\infty)
 \ena  
Hereafter,   $\psi_m=\psi(D/2)$ is 
  the order parameter  at the midpoint 
$z=D/2$ and   $f_m= f(\psi_m)$. From Eq.(2.13) 
$\omega-\omega_s$ is simply a constant.  We require   
$d\psi/dz =0$ at $z=D/2$ or $\omega=0$ at $\psi=\psi_m$ 
because of the  symmetric boundary conditions in 
 Eq.(2.22).  In the region $0<z<D/2$,  $\psi=\psi(z)$ 
 is obtained from 
 \be 
 D/2-z= \int_{\psi_m}^{\psi}d\psi \sqrt{C(\psi)/2\omega(\psi)}, 
\en 
where $\omega=\omega(\psi)$ 
is treated as a function of $\psi$. 
As $z\to 0$, the plate separation  distance $D$ is expressed as  
\be 
D= \int_{\psi_m}^{\psi_0}d\psi \sqrt{2C(\psi)/\omega(\psi)}, 
\en 
which  determines $\psi_m$ for each $D$, 
indicating the following.  
(i) In the integral of Eq.(2.26), 
we may push the upper bound $\psi_0$    
to infinity for large $\psi_0$ since 
$1/\sqrt{\omega} \sim \psi^{-(1+\delta)/2}$ 
for large  $\psi$. Thus  $\psi_m$ becomes  
independent of $\psi_0$ as $\psi_0\to \infty$. 
 (ii) For a thick film with  
  $D\gs  \xi_\infty$,  $\psi_m$  should approach   
 $\psi_\infty$, 
where $\xi_\infty$ is 
the correlation length in  Eq.(2.19). 
In the integrand of Eq.(2.26),  
we expand $\omega$ in powers of  
$\varphi \equiv \psi -\psi_m$ as 
\bea 
\omega&=& (\mu_m-\mu_\infty)
\varphi  + f''(\psi_m) \varphi^2/2 +\cdots,\nonumber\\
&\cong & f''(\psi_m) [ (\psi_m-\psi_\infty) 
\varphi + \varphi^2/2] ,
\ena 
 where 
$\mu_m=\mu(\psi_m)$. The  second line follows for 
$\psi_m\cong \psi_\infty$. 
Using  $\xi_\infty 
= [C(\psi_\infty)/f'' (\psi_\infty)]^{1/2}$, 
we perform   the integral 
in  Eq.(2.26) 
as $2\xi_\infty\ln [\psi_\infty/ (\psi_m-\psi_\infty)]$ 
for $\psi_\infty\neq 0$. 
Thus, for $D\gs \xi_\infty$, $\psi_m$ approaches $\psi_\infty$ as   
\be 
 \psi_m-\psi_\infty \sim \psi_\infty \exp({-D/2\xi_\infty}).
\en   

\begin{figure}[htbp]
\begin{center}
\includegraphics[scale=0.62]{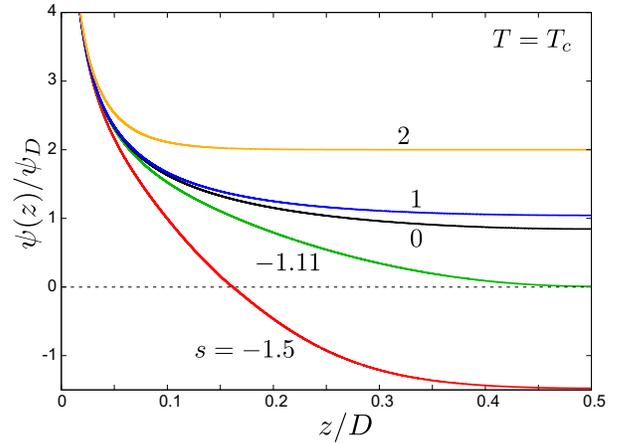}
\caption{\protect
(Color online) 
Normalized order parameter   
$\psi(z)/\psi_D$ 
vs $z/D$   at $T=T_c$ 
in a  film  in the half region  $0<z<D/2$ 
for $s=\psi_\infty/\psi_D= 2, 1,0, -1.11$, and -1.5  
from above  with $\psi_0=20 \psi_D$ in the symmetric boundary 
conditions, where $\psi_D$ is defined by Eq.(2.9).  
 The decay is algebraic for small $z$ as 
in Eq.(2.15).  
}
\end{center}
\end{figure}

\begin{figure}[t]
\begin{center}
\includegraphics[scale=0.44]{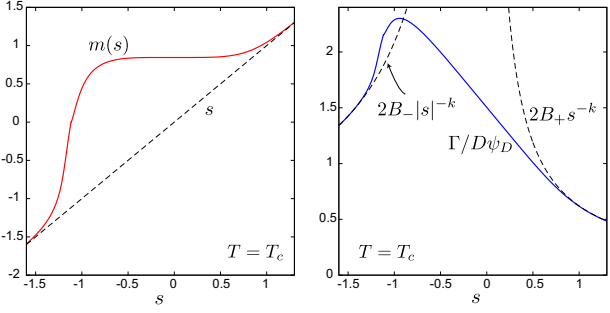}
\caption{\protect  
(Color online) 
Left: $m=\psi_m/\psi_D$ vs $s=\psi_\infty/\psi_D$ 
for a film at $T=T_c$ in the limit $\psi_0\to\infty$, 
where $\psi_m$ is the midpoint value, 
 $\psi_\infty$ is the reservoir value, and $\psi_D 
 \propto D^{-\beta/\nu}$. Here  $m$  stays at  $0.8$ 
in the weak response region ($-0.9 \ls s \ls 0.7$), 
changes steeply in the catastrophic region ($-1.2 \ls s \ls -0.9$), 
and tends to $s$ in the strong reservoir region 
($|s|\gs 1.1$).  
Right: Normalized adsorption $\Gamma/D\psi_D$ vs 
$s$ at $T=T_c$.
It approaches  $2B_+ s^{-k}$ for $s \gs 1$ 
and $2B_- |s|^{-k}$ for $s \ls -1$ 
(dotted lines) 
with $k= \nu/\beta-1$ from Eq.(2.20). 
The lines of $\Gamma/D\psi_D$ and 
$2B_- |s|^{-k}$ crosses at $s=-0.90$, 
where the Casimir amplitude 
$\Delta(s)$ is maximized from 
Eq.(2.52). 
}
\end{center}
\end{figure}  
  
In the limit $\psi_0 \to \infty$, 
the profile $\psi(z)$ is  scaled 
in terms of a scaling function $\Psi(u,s)$    
as 
\be 
\psi(z)=\psi_D \Psi(z/D, \psi_\infty/\psi_D).
\en 
Hereafter, we set  
\bea 
s&=& \psi_\infty/\psi_D , \\ 
m&=& \psi_m/\psi_D=\Psi(0.5,s),  
\ena 
which are   the normalized   
order parameter in the reservoir and that 
at the midpoint, respectively. 
In Fig.2, for $\psi_0/\psi_D=20$, we plot  
 $\psi(z) /\psi_D$ vs $z/D$  for five  values of $s$. 
These profiles 
represent $ \Psi(z/D,s)$  slightly away 
from the wall or  for $z\gg \ell_0$. 
For $s=2$ and $-1.5$, we find 
 $m\cong s$ (or $\psi_m\cong \psi_\infty$). 
However, the  curves  for $s=0$ 
and $s=1$ are  very close, while that for $s=-1.11$ 
tends to 0 at the midpoint.

\subsection{Weak response and catastrophic 
behaviors}
 
In the left panel of Fig.3, 
we show $m$ vs $s$ determined by Eq.(2.26) in the limit 
$\psi_0\to\infty$. The slope of the curve $dm/ds$ is written as 
\be 
 \frac{dm}{ds}= 
\ppp{\psi_m}{\psi_\infty}{D} =
\frac{\chi_m}{\chi_\infty}  ,
\en 
which is equal to  the ratio of the two susceptibilities 
$\chi_m= k_BT_c\p \psi_m/\p \mu_\infty$ at the midpoint and 
$\chi_\infty= 
k_BT_c \p \psi_\infty/\p \mu_\infty$ in the reservoir 
 \cite{Nakanishi-mean}. Here, 
  because of the large size of the critical 
 exponent $\delta=4.815$, 
 there are  distinctly different three 
 regions of $s$:     
(i) For $-0.9 \ls s \ls 0.7$, $m$ is nearly  a constant 
about 0.8  with  $dm/ds  \sim   0.1$. 
Here $m$ little changes 
 with a change in  $s$. The response of $\psi$ in 
the film to a change in  $\psi_\infty$ 
 is weak. 
(ii) For $-1.2 \ls s \ls - 0.9$, we have 
$dm/ds  \sim  5 \sim \delta$. 
In this catastrophic region, 
 $m$  changes steeply between   the 
base curve $m=s$  and   0.8. 
The width of this region is of order $1/\delta$ and is narrow. 
(iii) In the regions  $s \gs 1$ and $s \ls -1.3$, 
Eq.(2.28) gives 
\be 
m-s \sim 
\exp[{-\sqrt{\delta(\delta +1)/2}|s|^{\nu/\beta}}]. 
\en 
That is,  $\psi_m \cong \psi_\infty$ and 
the reservoir inflence is strong.

The excess adsorption $\Gamma=\int_0^D dz (\psi-\psi_\infty)$ 
in the film with respect to the reservoir 
 is expressed as    
\be
\Gamma 
=   \int_{\psi_m}^{\psi_0}
 d\psi  (\psi-\psi_\infty) \sqrt{{2C(\psi)}/{\omega(\psi)}} 
\en
Again we may push the upper bound of the integral  
to infinity for large $\psi_0$; then, 
   $\Gamma/D\psi_D$ becomes a universal function of $s$. 
   At $T=T_c$,  Eq.(2.20) 
  indicates that   $\Gamma \to 
 2\Gamma_{{\rm sem}\pm}=2D\psi_D  B_\pm |s|^{1-\nu/\beta}$ 
for $D \gs \xi_\infty$ or for $|s|\gs 1$.  
 It is convenient 
to introduce a scaling function $\Gamma^*(s)$ by 
\be 
\Gamma =
 2\Gamma_{{\rm sem} \pm} +D\psi_D \Gamma^*.
\en   
In the right panel of 
Fig.3, we show $\Gamma/D\psi_D$ vs $s$ in the limit 
$\psi_0\to\infty$. 
In the weak response region, $-0.9 \ls s \ls 0.7$,  
we have $\Gamma/D\psi_D \sim  0.5-  s $.   
Thus   $\Gamma/D\psi_D$ 
increases  with decreasing $s$,  
exhibiting a peak  
at the border of the weak response 
and catastrophic regions. In fact, 
 its maximum is 2.29 at $s=-0.94$.  
In the strong reservoir region, 
 it  approaches 
  $2B_\pm |s|^{-k}$ with $k=\nu/\beta-1$ 
   (see the dotted lines 
  in the right panel of Fig.3).

\subsection{Casimir term in the force density}

Using Eq.(2.23),  
the excess grand potential $\Omega$ in Eq.(2.21) 
is expressed in terms of $\omega=\omega(\psi)$   as 
\bea 
\Omega &=&  D [f_m-f_\infty- 
 \mu_\infty (\psi_m-\psi_\infty)] \nonumber\\
&& +  2 \int_{\psi_m}^{\psi_0} 
d\psi  \sqrt{2C(\psi) \omega(\psi)}.
\ena  
In the right hand side, the first term arises due to 
 the reservoir. 
In the second  term, the gradient contribution 
gives rise to the factor 2.
Let us calculate  the derivative  
$ \p \Omega/\p D$ 
at fixed $\psi_0$  and  $\psi_\infty$, treating 
 $\psi_m$ as  a function of  $D$. 
  The derivative of the 
second  term is $D (\mu_\infty -\mu_m)\p \psi_m/\p D$  from 
Eq.(2.26), where 
$\mu_m=f'(\psi_m)$. 
 We then find a simple expression, 
\be 
\frac{\p\Omega}{\p D}   = f_m-f_\infty- 
 \mu_\infty (\psi_m-\psi_\infty). 
\en 
The Casimir amplitude of the force density 
${\cal A} = ({\p\Omega}/{\p D}){D^d }/{k_BT_c}$ 
is expressed as   
\be
{\cal A} = \frac{D^d }{k_BT_c} \bigg[f_m-f_\infty- 
 \mu_\infty (\psi_m-\psi_\infty)\bigg].
\en
Note that the osmotic pressure $\Pi$ 
 is the force density per unit area 
 exerted by the fluid  to the walls. See Appendix A for more discussions. 
 Thus we find \cite{Tsori}   
\be 
\Pi=-\frac{\p \Omega}{\p D} = 
-k_B T_c {\cal A}/D^d. 
\en 
From the second line of Eq.(2.24) 
we also notice the relation 
$\omega(\psi)-\omega_s(\psi) 
 = -\p\Omega/\p D$. 
Equations (2.36)-(2.39) 
 hold   even for 
$T\neq T_c$ in our theory  
 in the next section.

At $T=T_c$,  use  of Eqs.(2.3), (2.8),  and (2.9) gives   
\be 
\frac{1}{ A_c} {\cal A}(s)=\delta|s|^{1+\delta} 
+ |{m}|^{1+\delta}
 -(1+\delta)|s|^{1+\delta}\frac{m}{s}.  
\en 
in terms of $s$ in  Eq.(2.30)  and $m$ in  Eq.(2.31).   
We can see that 
${\cal A}={\cal A}(s)$ is a universal function 
of $s$ as  $\psi_0\to \infty$.   
In Eq.(2.26) we set $\psi= \psi_Dq$ to obtain  
\be 
\int_{m}^\infty dq  q^{-\eta\nu/2\beta}
/\sqrt{{{\hat{\omega}}}(q,s,m) }=1,
\en 
where   $\hat{\omega}= \omega/B_0\psi_D^{1+\delta}$   depends   on 
 $q$, $s$, and $m$  as  
\be 
\hat{\omega}= 
|q|^{1+\delta}- |m|^{1+\delta} 
- (1+\delta)|s|^{\delta-1}s (q- m).
\en 
We  seek $m=m(s)$ for each  $s$ from Eq.(2.41) 
(as given  in  the left panel of Fig.3).
 In particular, 
for  $|s| \gg 1$, we have  $\psi_m\to \psi_\infty$ and  
$\p \Omega/\p D \propto (\psi_m-\psi_\infty)^2$ 
from Eq.(2.37). From Eq.(2.28)   ${\cal A}(s) $ 
decays  for $|s|\gs 1$ as   
\be
{\cal A}(s) \sim  e^{-D/\xi_\infty}  
\sim  \exp[{-\sqrt{2\delta(\delta +1)}|s|^{\nu/\beta}}].
\en

We calculate  ${\cal A}(s)$  
for two special    cases.  
(i) First, for $s=0$, 
the reservoir is at the criticality. 
Here,  $m>0$, so  
setting $v=q/m$ in  Eq.(2.41) gives    
 $m^{\nu/\beta}=I_0$ with 
\be 
I_0= \int_1^{\infty} dv {v^{-\eta\nu/2\beta}
}/{\sqrt{v^{1+\delta}-1}}.  
\en 
where $I_0=0.719$ for $d=3$. Since  
 $\p\Omega/\p D= f(\psi_m)$,  we  obtain 
the critical-point  value ${\cal A}_{\rm cri}= {\cal A}(0)$ 
in the form,  
\be 
{\cal A}_{\rm cri} =   I_0^d  A_c ,
\en  
where $A_c$ is given by Eq.(2.8) 
and will be estimated below Eq.(3.16). 
Essentially the same calculation 
was originally due to Borjan and Upton  \cite{Upton}.  
(ii) Second, we assume   $\psi_m=0$, which is attained 
for $\psi_\infty<0$ or for  $s<0$.  See the corresponding 
curve of $s=-1.11$ in Fig.2. From Eq.(2.41) 
we  obtain  $|s|^{\nu/\beta} = J_0$ with 
\be 
J_0= \int_0^{\infty} dv v^{-\eta\nu/2\beta}
/\sqrt{v^{1+\delta}+(1+\delta)v },
\en 
where   $J_0= 1.195$ at $d=3$ numerically, 
leading to $s=- J_0^{\beta/\nu}= -1.1170$. 
 From  Eqs.(2.40) and (2.45)  we find    
\be 
{\cal A}(- 1.1170) =  J_0^d A_c  \delta=
{\cal A}_{\rm cri}  (J_0/I_0)^d\delta ,
\en  
Thus  ${\cal A}(- 1.1170)/{\cal A}_{\rm cri}=24.45$ for $d=3$.  

In Fig.4, 
 we display  
${\cal A}(s)/{\cal A}_{\rm cri}$ 
for $d=3$ calculated from 
Eqs.(2.40) and (2.41). 
Its maximum is  24.45  at $s=-1.1170$,  where 
 $\psi_m\cong 0$. To be precise,   the 
curve exhibits a small  cusp  due to  
the weak $\psi$-dependence of $C(\psi)$.     
Similar enhancement of 
$\cal A$ was found at off-critical compositions by 
Macio\l ek {\it et al} \cite{Evans-Anna} 
for two-diemnsional Ising films 
and by Schlesener {\it et al.} \cite{JSP}  
in the mean-field theory at $T=T_c$. 
  In the next section, the origin of  
this  peak will be ascribed to  the fact that 
the peak point on the line $T=T_c$ is 
  very close to  
a capillary-condensation critical point 
 in the region  $T<T_c$ 
(see Figs.9 and 12).

Mathematically, the  peak of  ${\cal A}(s)$ at $T=T_c$ 
stems  from the presence of the  
weak response and catastrophic regions, for which see the 
explanation of the right panel of Fig.3. 
Here 
we calculate $d{\cal A}(s)/ds$ from  Eq.(2.40)  as 
\be
\frac{1}{a|s|^{\delta -1}}\frac{d{\cal A}}{ds} =  
\bigg ({\bigg|\frac{m}{s}\bigg|^{\delta-1}}\frac{m}{\delta} -
\frac{s}{\delta} \bigg)\frac{dm}{ds}- m+s, 
\en   
where $a= A_c(1+\delta) \delta \sim 28A_c\sim 40$.   
In the weak response region, 
we may neglect the first term in the right hand side  
to obtain 
$d{\cal A}/ds \cong 
- a |s|^{\delta -1}(m-s)<0$, 
which is nearly zero for $s\gs -0.5$ 
and grows abruptly for $s\ls -0.5$. 
In the catastrophic region, 
$dm/ds$ is of order $\delta$  such that  
$d{\cal A}/ds$ changes its sign, leading to 
a maximum of $\cal A$.  

\subsection{Casimir term in the grand potential}

For binary mixtures, the   de Gennes-Fisher  scaling form 
\cite{Fisher} for the grand potential reads  
\be 
\Omega= \Omega_\infty  - k_BT D^{-(d-1)} {\Delta}, 
\en 
where  
$\Omega_{\infty}=\lim_{D\to \infty}\Omega$  
is the large-separation limit. The  $\Delta$ is a function 
of  $s$ in Eq.(2.30) (and a scaled reduced 
temperature in the next section). 
This form  may be  inferred from  
the boundary behavior of  
 $f_{\rm loc}$ in Eq.(2.17). 
Next, we  differentiate   Eq.(2.36) 
with respect to  $\mu_\infty$ 
(or   $\psi_\infty$)  at fixed $D$ and $\psi_0$.  
Following the  
procedure used in deriving  Eq.(2.37), we  obtain 
the Gibbs adsorption formula \cite{Evans-Marconi}, 
\be 
\frac{\p \Omega}{\p\mu_\infty}= 
\frac{\chi_\infty}{k_BT_c}  
\frac{\p \Omega}{\p\psi_\infty}= 
- \Gamma,
\en 
where $\Gamma$ is the excess adsorption in 
Eq.(2.34) and 
$\chi_\infty= k_BT_c/f''(\psi_\infty)$. If  
 the Fisher-de Gennes form (2.49) is substituted into 
 Eq.(2.50), 
the  above relation yields   
\be 
\frac{\p \Delta}{\p s}=R_\infty   \Gamma^*
\en 
where $\Gamma^*$ is defined by Eq.(2.35) 
and $R_\infty$ is the following 
dimensionless combination representing 
the scaled inverse susceptibility in the reservoir, 
\be 
R_\infty= {D^d} \psi_D^2/\chi_\infty.
\en 
At $T=T_c$, we obtain 
$R_\infty= A_c \delta(\delta+1) |s|^{\delta-1}$, 
which has already appeared in Eq.(2.48)  
as $a|s|^{\delta-1}$.   In our theory 
 in the next section,
  Eqs. (2.49)-(2.52) will  remain valid 
  even for $T\neq T_c$. 

At $T=T_c$,  direct differentiation of $\Omega$ 
in Eq.(2.49) with respect to $D$ 
 also yields a relation 
between  ${\cal A}(s)$ and $\Delta (s)$, 
\be 
{\cal A}(s)= 
(d-1){ \Delta}(s)
 -\frac{\beta s}{\nu} \frac{\p  }{\p s} {\Delta}(s).   
\en 
Elimination of $\p \Delta/\p s$ from Eqs.(2.51) and (2.53) 
yields      
\be 
\Delta (s) =
  \frac{{\cal A}(s)}{d-1} 
+  \frac{d  A_c }{d-1}\delta 
|s|^{\delta-1}s\Gamma^*(s). 
\en
Thus, as well as ${\cal A}$, $\Delta $ is proportional to 
the universal number 
$ A_c$. The critical-point  values 
  of the amplitudes, written as  
 $\Delta_{\rm cri}=\Delta(0)$ 
and  ${\cal A}_{\rm cri}={\cal A}(0)$,  are  related as  
\be 
\Delta_{\rm cri} = {\cal A}_{\rm cri}/(d-1) .
\en

In Fig.4, we  also plot 
${\Delta}(s)/{\Delta}_{\rm cri}$ for $d=3$ 
numerically calculated from 
Eq.(2.54). Its peak height is 
 $13.36$ at $s=-0.90$, which is about half of the 
height of ${\cal A}(s)/{\cal A}_{\rm cri}$.   
The differential equation (2.53) is excellently 
satisfied by   ${\cal A}(s)$ 
and ${\Delta}(s)$ in Fig.4. 
From Eq.(2.51) the amplitude 
$\Delta(s)$ is maximized at a point 
where  $\Gamma^*=0$. 
Indeed, in the right panel of Fig.3,  
the line of $\Gamma/D\psi_D$ vs $s$ 
and that of $2B_{-}|s|^{\nu/\beta-1}$ vs $s$ cross   
at $s=-0.90$, where $\Delta (s)$ is maximum in Fig.4. 


\begin{figure}[t]
\begin{center}
\includegraphics[scale=0.62]{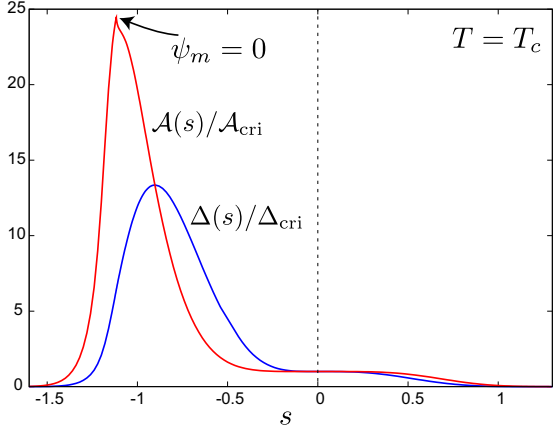}
\caption{\protect  
(Color online) 
Casimir amplitude ratios 
 ${\cal{A}}(s)/{\cal{A}}_{\rm cri}$ for the force 
 density  and  $\Delta(s)/\Delta_{\rm cri}$ for the grand potential    
vs $s=\psi_\infty/\psi_D$ at $T=T_c$ 
in the limit $\psi_0\to\infty$. 
These quantities are maximized at $s\sim -1$ upon the changeover 
between the weak response and catastrophic regions (see 
the text).  
}
\end{center}
\end{figure}

\section{Critical behavior for $T \neq T_c$}
\setcounter{equation}{0}
In this section, we now set up a local functional theory  
including the gradient free energy for nonvanishing 
reduced temperature,  
\be 
\tau=T/ T_c-1.
\en  
For binary mixtures, we suppose  the 
upper  critical solution temperature (UCST), 
while $\tau$  should be defined as  $\tau =1-T/T_c$ 
for the lower critical solution temperature (LCST).  
Our model is similar  to 
the linear parametric model by Schofield {\it et al.}
\cite{Sc69,Hohenberg,Wallace2} 
and the local functional model 
by  Upton {\it et al.} 
\cite{Fisher-Upton,Upton,Upton1,Upton-ad}. 
 The latter is composed of     the free energy  
of the linear parametric model    and the 
gradient free energy.  We use a simpler 
 free energy 
accounting for the renormalization effect 
and the gradient free energy. 
Further, we define  the free energy  
 within the coexistence curve 
and make it  satisfy the two-scale-factor 
universality\cite{Stauffer,HAHS}. 
 Appendix B  will give   the relationship 
between  our model and the linear parametric  model.

The  Casimir amplitudes  
${\cal A}$  in Eq.(2.38)  and ${\Delta}$ 
in  Eq.(2.49) depend  on $s=\psi_\infty/\psi_D$ 
in Eq.(2.30) and the scaled reduced temperature, 
\be 
t= \tau (D/\xi_0)^{1/\nu}.
\en 
where $\xi_0$ is a  microscopic  length ($\sim 
3 \rm{\AA}$)  in the   correlation length 
$\xi=\xi_0 \tau^{-\nu}$ for 
$\tau>0$ at the critical composition.  
For example,  $\xi/D$ is $0.24$ for $t=10$ 
and $0.056$ for $t=100$.

\subsection{Model outside the coexistence curve}

For $\tau<0$,  two phases 
can  coexist  
in the bulk. The coexistence curve 
is written as  $\psi=\pm \psi_{\rm cx}$ with 
\be 
\psi_{\rm cx} = b_{\rm cx}|\tau|^\beta,
\en  
where  $ b_{\rm cx}$ is a constant. 
In this subsection, 
we present a  local free energy 
density $f_{\rm loc}$ applicable outside the coexistence curve 
($|\psi|\ge \psi_{\rm cx}$ if  $\tau<0$). 
In the next subsection, we present  its 
form within  the coexistence curve.

As a generalization of the model 
in Eqs.(2.1)-(2.4) for $\tau=0$, 
we again use  the usual form,  
\be
f_{\rm loc} =f + \frac{1}{2}k_BT_c C|\nabla\psi|^2.
\en 
The free energy density 
 $f=f(\psi,\tau)$ is an even function of  $\psi$ 
expressed as 
\be 
f =  k_BT_c\bigg (
\frac{1}{2}r\psi^2+ \frac{1}{4}u\psi^4 \bigg).  	
\en 
Here  $r$, $u$, and $C$ are 
 renormalized coefficients depending 
 on $\psi$ and $\tau$. 
To account for the renormalization 
effect, we introduce a distance from the critical point $w$.  
In the  critical region ($w \ll 1$), these 
coefficients   depend on $w$ as 
\bea 
C& =& C_1 w^{-\eta\nu},  \\
r&=& C_1 \xi_0^{-2}w^{\gamma-1} \tau ,\\
u&=&  C_1^2u^* \xi_0^{-\epsilon}w^{(\epsilon-2\eta)\nu} ,
\ena
where $C_1$ and  $u^*$ are   constants. 
For $\tau>0$ and $\psi=0$, 
we set $w=\tau$ to obtain $\xi^2=C/r$. 
For $\tau=0$,   
 we  should have 
$w^{2\beta} \propto  \psi^{2}$. 
We thus determine  $w$ as a function of $\tau$ and 
$\psi^2$ from 
\be 
w=  \tau +C_2 w^{1-2\beta}\psi^2,    
\en 
where $C_2$ is a  constant. 
Recall   the mean-field 
expression $\chi^{-1}=r+3u\psi^2$ 
for the susceptibility $\chi$, 
which holds even for $r<0$.  
If we use Eqs.(3.6)-(3.8), 
we have 
$r+3u\psi^2= C_1\xi_0^{-2}w^{\gamma-1}
[\tau + 3C_1 u^* \xi_0^{2-\epsilon}w^{1-2\beta}\psi^2]$ 
in our renormalized theory.  
Thus we may require 
\be 
r+3u\psi^2 = C_1\xi_0^{-2}w^{\gamma},
\en  
where 
 we relate $C_2$ in Eq.(3.9) to $C_1$   as 
\be 
C_2=       3u^* C_1\xi_0^{2-\epsilon}.  
\en 
See Fig.5 for $f$ and $w$ as functions of  
$\psi$ for fixed positive $\tau$. 
In the simplest case  $\tau>0$ and $\psi=0$, 
we have $w= \tau$ and $w^\nu=\xi_0/\xi$ so that the 
concentration susceptibility $\chi(\tau,\psi) 
= k_BT_c(\p \psi/\p \mu)_\tau$  
 grows strongly with the  exponent $\gamma$ as      
\be 
\chi(\tau,0) = r=  C_1^{-1}\xi_0^{2} \tau^{-\gamma}. 
\en

In  the renormalization 
group theory \cite{Onukibook}, Eqs.(3.6)-(3.9) 
follow if $\xi_0^{-1}w^\nu$ 
is the lower cut-off wave number 
of the renormalization. The coupling constant 
$u^*$ should be a universal fixed-point value. 
Its  $\epsilon$ expansion reads   
\be 
K_d u^* =\epsilon/9 +\cdots, 
\en   
where $K_d$ is the surface area of a unit sphere 
 in $d$ dimensions 
divided by $(2\pi)^d$ (so $K_4= 1/8\pi^2$ and 
 $K_3= 1/2\pi^2$).  
Retaining the small critical 
exponent $\eta \propto\epsilon^2$ and using the relations 
among the critical exponents, 
we may    correctly describe 
 the  asymptotic scaling  behavior, though  the 
 critical amplitude ratios are approximate. 
In addition, note that   a constant 
term independent of $\psi$ has been omitted 
in the singular free energy density in Eq.(3.5), 
which yields the singular specific heat\cite{constant}. 

For $\tau=0$, 
the  Fisher-Au Yang 
model in Eq.(2.1) folows 
with $C_0$ and $B_0$ given by    
\bea 
C_0/k_BT_c & =& C_1 C_2^{-\eta\nu/2\beta},\\
B_0/k_BT_c & =& u^* C_1^2 C_2 ^{(\epsilon/2-\eta)
\nu/\beta} \xi_0^{-\epsilon}/4. 
\ena 
Thus  we have $B_0k_BT_c/C_0^2= 
 u^*  C_2 ^{\epsilon\nu/2\beta} 
 /4\xi_0^{\epsilon}$ and 
$ B_0/C_0  = u^* C_1 C_2 ^{\nu/\beta-1} 
\xi_0^{-\epsilon}/4 = 
C_2 ^{\nu/\beta} /12\xi_0^2$.  
The  universal number $A_c$ 
in Eq.(2.8) is calculated  as 
\be
A_c 
=  2^d 6^{-\epsilon/2}/u^*. 
\en   
As a rough estimate, 
we set $u^*= 1/9K_3= 
2\pi^2/9$ for $d=3$ from Eq.(3.13). This  
leads  to 
$A_c=(2^3/\sqrt{6})/u^*= 1.49$. 
This value yields 
$\Delta_{\rm cri}= {\cal A}_{\rm cri}/2= 
0.279$ from Eq.(2.45).
For $d=3$, 
Krech  \cite{Krech} estimated 
$\Delta_{\rm cri}$ to be  $ 0.326$ by a field-theoretical 
methods and $ 0.345$ 
by a Monte Carlo method, 
Borjan and Upton  \cite{Upton} 
obtained $\Delta_{\rm cri}=0.428$ 
by the local functional theory, and 
Vasilyev {\it et al.} \cite{Monte}  
found   ${\Delta}_{\rm cri}=0.442$ 
by a Monte Carlo  method.  

To make the following expressions for $f$ and $\mu$ simpler, 
we introduce the dimensionless  ratio, 
\be 
S=\tau/w, 
\en 
in terms of which $\psi^2$ is expressed as 
\be 
\psi^2= (1-S)w^{2\beta}/C_2 .
\en 
Then  $f$ and   $\mu=( \p f/\p \psi)_\tau$ 
 are expressed   as 
\bea 
&&\hspace{-1cm}
\frac{f}{k_B T_c}=
\frac{w^{2-\alpha}}{36u^*\xi_0^{d}}(1+5S)(1-S) ,
\\ 
&&\hspace{-1cm}
\frac{\mu}{ k_BT_c}= 
  \frac{2-\alpha+ 4(1-\alpha)S +5\alpha S^2 }{18u^*[2\beta 
 +(1-2\beta)S]\xi_0^{d}}w^{\gamma}
  C_2\psi.  
\ena  
 In  Eq.(3.20), 
we have used  the relation   
$({\p w}/{\p \psi})_{\tau}=
{2C_2w^{1-2\beta}\psi}/[{ 2\beta 
 +(1-2\beta)S} ]$. 
However, the susceptibility 
$\chi(\tau,\psi)= k_BT_c/(\p \mu/\p \psi)$ 
is somewhat complicated  \cite{sus}. It can be 
simply calculated for $\mu=0$ as in  Eq.(3.12) for $\tau>0$  
and  as in Eq.(3.24) below for $\tau<0$.

We  now seek  the coexistence curve (3.3) by  setting   
 $\mu=0$ with $\tau<0$. 
From Eq.(3.20) it follows the quadratic equation  
 $2-\alpha+ 4(1-\alpha)S +5\alpha S^2 =0$ of $S$, 
which is solved to give    $S=-1/\sigma$ or 
\be 
 w= \sigma |\tau|. 
\en   
The coefficient  $\sigma$ is  expressed in terms of $\alpha$ as 
\be
\sigma=2 - 9\alpha/[ 2 + \sqrt{4-18\alpha+9\alpha^2}] ,
\en  
which is equal to 1.714 for $d=3$. 
If we substitute Eq.(3.21) into  Eq.(3.9), 
the coefficient  $b_{\rm cx}$ in Eq.(3.2) 
is calculated    as 
\be 
b_{\rm cx}= C_2^{-1/2} (1+\sigma)^{1/2}\sigma^{\beta-1/2},    
\en 
leading to  $C_2=  (1+\sigma)\sigma^{2\beta-1}/b_{\rm cx}^2$. 
Since $b_{\rm cx}$ is experimentally measurable,  
there remains no arbitrary parameter with Eqs.(3.11) and (3.23).  
The susceptibility $\chi$  on the coexistence curve is expressed as 
\be 
\chi(-|\tau|, \psi_{\rm cx})= 
\frac{3\sigma}{2\zeta(1+\sigma)}C_1^{-1}\xi_0^2 |\sigma\tau|^{-\gamma}.
\en 
In  the denominator of this relation, we introduce
\be 
\zeta= \frac{2(1-\alpha)\sigma-5\alpha}{
(2\beta\sigma-1+2\beta)^2} ,
\en 
which   is equal to 4.28 for $d=3$. 
The ratio of the  susceptibility  for $\tau>0$ and $\psi=0$ 
and that on the coexistence curve at the same $|\tau|$ is written as 
\be 
R_\chi= \frac{\chi(|\tau|,0)}{
\chi(-|\tau|, \psi_\infty)} =
\frac{2}{3}\zeta (1+\sigma)\sigma^{\gamma-1},
\en 
which is  8.82 for $d=3$. Note that  
the   $\epsilon $ expansion  gives 
$R_\chi= (2+\epsilon)2^{\epsilon/6}+\cdots$ 
and its reliable estimate is 4.9  
\cite{Nicoll,Liu1,Onukibook}.  We also write 
the correlation length on the coexistence 
curve as   $\xi=(C/f'')^{1/2}= 
\xi_0' |\tau|^{-\nu}$, where $\xi_0'$ is another microscopic length. 
The ratio of the two microscopic 
lengths $\xi_0$ and $\xi_0'$ is written as  
\be 
\xi_0/\xi_0'=[2\zeta (1+\sigma)\sigma^{2\nu-1} /3]^{1/2}, 
\en   
which gives  $\xi_0/\xi_0'= 2.99$ for $d=3$. 
Note that the $\epsilon $ expansion result is 
$\xi_0/\xi_0'= 2^{\nu}(1+ 5\epsilon/24+\cdots)$ 
and its reliable estimate is 1.9  
for $d=3$ \cite{Nicoll,Liu1,Onukibook}. 
  We recognize that the correlation length $\xi$ 
and the susceptibility $\chi$  on the coexistence curve 
are considerably underestimated in our theory (mainly due to 
a factor $Z(\theta)$ in Eq.(B13) in Appendix B).

Finally, let us consider the  characteristic order parameter 
of a film $\psi_D$ defined in Eq.(2.9).  From the sentence below 
Eq.(3.15) it  is written as   
\bea
\psi_D &=& C_2^{-1/2} (
\sqrt{24}\xi_0/D)^{\beta/\nu} \nonumber\\
& =&  1.47 b_{\rm cx}(\xi_0/D)^{\beta/\nu}  
\ena 
where $b_{\rm cx}$ is given by  Eq.(3.23).
  Equation (3.18) gives  an 
 expression for $\psi$ valid outside the 
 coexistence curve, 
\be 
|\psi|/\psi_D= 24^{-\beta/2\nu} (1-S)^{1/2}|S|^{-\beta} 
|t|^{\beta}  .
\en 
On the coexistence curve, this  expression becomes  
\be 
{\psi_{\rm cx}}/{\psi_D}  = 
24^{-\beta/2\nu}(1+\sigma^{-1})^{1/2}  \sigma^{\beta}   
|t|^\beta, 
\en 
which is equal to $0.66 |t|^\beta$ in our theory.

\begin{figure}[t]
\begin{center}
\includegraphics[scale=0.43]{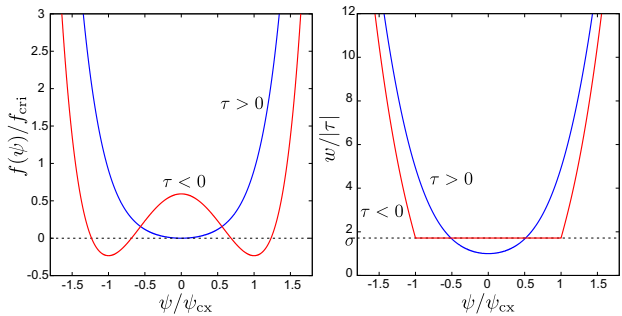}
\caption{\protect  
(Color online) 
Normalized singular free energy density 
 $f(\psi)/f_{\rm cri}$ (left) and 
 normalized distance from 
 the criticality $w/|\tau|=1/|S|$ (right) 
 vs $\psi/\psi_{\rm cx}$, 
 where $f_{\rm cri}= k_BT_c|\tau|^{2-\alpha}/u^* \xi_0^d$ 
 and $\psi_{\rm cx}=b_{\rm cx}|\tau|^\beta$ 
 (both for $\tau>0$ and $\tau<0$). The upper 
 (lower) curves correspond to those for $\tau>0$ ($\tau<0$). 
 For $\tau<0$, $w$ is independent of $\psi$ within 
 the coexistence curve $|\psi|/\psi_{\rm cx}<1$ in our model.}
\end{center}
\end{figure}  

\begin{figure}[htbp]
\begin{center}
\includegraphics[scale=0.62]{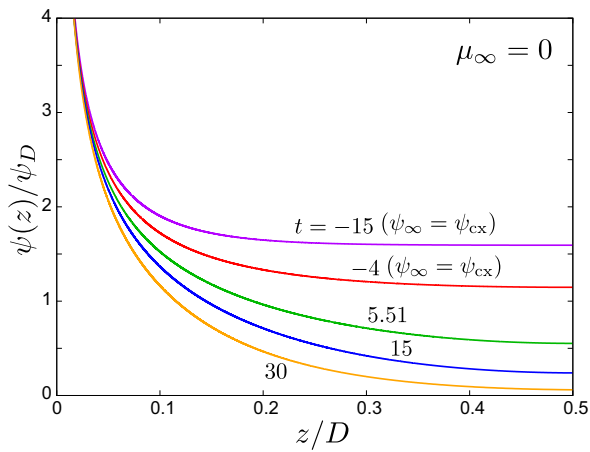}
\caption{\protect 
(Color online) 
Normalized order parameter   
$\psi(z)/\psi_D$ 
vs $z/D$   for $\mu_\infty=0$ 
in a  film  in the half region  $0<z<D/2$ 
for $t= 30, 15, 5.51$,  -4, and -15   
from below with $\psi_0=20 \psi_D$ in the symmetric boundary 
conditions. Here $\psi_\infty=0$ for $t>0$ and 
 $\psi_\infty=\psi_{\rm cx}$ for $t<0$. 
 The decay is algebraic for small $z$ as 
in Eq.(2.15).  
}
\end{center}
\end{figure}

\subsection{Model including  the coexistence curve interior}

For $\tau<0$, we need to 
define  a local free energy 
density $f$  inside the coexistence curve $|\psi|<\psi_{\rm cx}$, 
where the fluid is metastable or unstable in the bulk. 
Notice that  $\psi(z)$ changes from $-\psi_{\rm cx}$ 
to  $\psi_{\rm cx}$ in  the interface   region 
 in two-phase coexistence.  Thus, to calculate the surface 
tension in the Ginzburg-Landau scheme, 
we need $f(\psi)$ in  the region 
$|\psi|\le \psi_{\rm cx}$  for $\tau<0$. 
Since the linear parametric model 
is not well defined within  the coexistence 
curve, Fisher {\it et al.}
 \cite{Fisher-Upton} proposed 
its generalized form   
applicable  even within the coexistence curve 
to obtain  analytically continued 
van der Waals loops.  We propose a  much 
simpler model, though  the derivatives  $\p^\ell f/\p\psi^\ell$ 
  in our model are continuous 
only up to $\ell=2$  on the coexistence curve.

Within the coexistence curve, we assume a 
 $\psi^4$ theory including the gradient free 
 energy.  Since $\mu=\p f/\p \psi$ vanishes on the coexistence 
curve, $f$ is of the form,     
\be 
{f} ={f_{\rm cx}}
+\frac{1}{4}{ k_BT_c}  B_{\rm cx}(\psi^2 -\psi_{\rm cx}^2)^2,   	
\en 
where ${f_{\rm cx}}$ is the value of $f$ on the coexistence 
curve. From  Eq.(3.19) it is written as   
\be
f_{\rm cx} ={k_B T_c}
(\sigma-5)
(1+\sigma)|\tau|^{2-\alpha} 
/{36u^*\xi_0^{d}\sigma^\alpha}.
\en 
We determine  the coefficient ${B_{\rm cx}}$ 
 requiring  the continuity 
of the second derivative $f''=k_BT_c/\chi$. 
Then we obtain $2B_{\rm cx}\psi_{\rm cx}^2= 1/\chi$, 
where $\chi$ is given by Eq.(3.24). Some calculations give  
a simple result,
\be 
B_{\rm cx}= \zeta u_{\rm cx},
\en 
where $\zeta$ is given by Eq.(3.25) 
and  $u_{\rm cx}$ is   the value of $u$ in Eq.(3.8) 
on the coexistence curve written as 
\be 
u_{\rm cx}=  C_{1}^2u^* \xi_0^{-\epsilon} 
|\sigma\tau|^{(\epsilon-2\eta)\nu}.
\en 
It  follows the relation  
 $f_{\rm cx}= {k_B T_c}u_{\rm cx}\psi_{\rm cx}^4 
({\sigma-5})/{4(1+\sigma)}$.
The coefficient of the gradient term  
is  replaced by its value 
$C_{\rm cx}$ on the coexistence curve:    
\be
C_{\rm cx} = C_1 |\sigma\tau|^{-\eta\nu},  
\en 
Then the susceptibility $\chi$ and the correlation length $\xi$ 
are continuous across the coexistence curve.  
 In this model, the renormalization effect 
inside the coexistence curve 
is  the same as that on the 
coexistence curve at the same $\tau$. 
See Fig.5 for $f$ and $w$ vs  $\psi$ for fixed 
negative  $\tau$.

The surface tension $\sigma_s$ is given by 
 the classical formula 
$ 
\sigma_s= 2
k_BT_c C_{\rm cx} \psi_{\rm cx}^2/3\xi$  
 in   the $\phi^4$ theory \cite{Onukibook}. It is proportional to 
$\xi^{-d+1}$ with $\xi=\xi_0' |\tau|^{-\nu}$. 
Some calculations give the universal number, 
\bea 
A_s &=&  \sigma_s \xi^{d-1}/k_BT_c\nonumber \\
&=& 
[2\zeta(1+\sigma^{-1})/3]^{\epsilon/2}
/3\zeta u^*.
\ena 
Using  Eq.(3.25), we have $A_s= 0.165/u^* $. 
Further,  if we set $u^*= 2\pi^2/9$, 
we obtain $A_s =0.075$. 
Note that $A_\sigma$ is known to be about 0.09 for 
Ising-like systems  
\cite{Moldover}.

\subsection{Casimir amplitudes   for $T \neq T_c$}

From  Eq.(2.38) we can readily calculate ${\cal A}(t,s)$ 
numerically as a function of $s$ and $t$. 
However, to  calculate  $\Delta(t,s)$,  
we cannot use Eq.(2.54)  for $t\neq 0$ 
and need to devise  another 
expression.  To this end, we  write  the second term 
in the right hand of  Eq.(2.36)  as 
\be 
J=\Omega -D\frac{\p\Omega}{\p D}
=2 \int_{\psi_m}^{\psi_0}d\psi 
\sqrt{2C(\psi)\omega(\psi)},
\en  
where $\omega(\psi)$ is defined  by  Eq.(2.24) with $f$ 
being given by Eqs.(3.5) and (3.31). 
In the limit $D\to \infty$, we have 
$J \to \Omega_\infty$, where $\Omega_\infty$ 
is the large-separation limit:   
\be 
\Omega_\infty= 2 \int_{\psi_\infty}^{\psi_0}d\psi 
\sqrt{2C(\psi)\omega_s (\psi)}.   
\en 
Here,  the lower bound of the 
integration is $\psi_\infty$ and 
$\omega_s(\psi)$ is the grand potential density 
for the semi-infinite case in the form of  Eq.(2.13). 
Dividing the integration region in Eq.(3.38) 
into $[\psi_m,\psi_0]$ and  $[\psi_\infty,\psi_m]$,  
we find 
\bea 
J-\Omega_\infty&=& 2 \int_{\psi_m}^{\psi_0}d\psi 
{\sqrt{2C(\psi)}}[\sqrt{\omega (\psi)}- 
\sqrt{\omega_s (\psi)}]  \nonumber\\
&&\hspace{-0.5cm} - 2 \int_{\psi_\infty}^{\psi_m}d\psi 
\sqrt{2C(\psi)\omega_s (\psi)}.  
\ena 
In the first integral, 
 we may push the upper bound $\psi_0$ to infinity, 
 since the integrand tends to zero rapidly 
 for large $\psi$  
from  $\sqrt{\omega }-
\sqrt{\omega_s}= - (\p\Omega/\p D)/[\sqrt{\omega}+
\sqrt{\omega_s }]$ (see the sentence below Eq.(2.39)). 
From Eqs.(2.37) and (2.49) we obtain 
\be 
\Delta=- {\cal A} - \frac{D^{d-1}}{k_BT_c} 
(J-\Omega_\infty).  
\en 
With this expression,  we can  calculate  
 $\Delta(s,t)$  numerically, We  
  confirm that  it is  a function of  
$s$ and $t$ only.  

\begin{figure}[t]
\begin{center}
\includegraphics[scale=0.43]{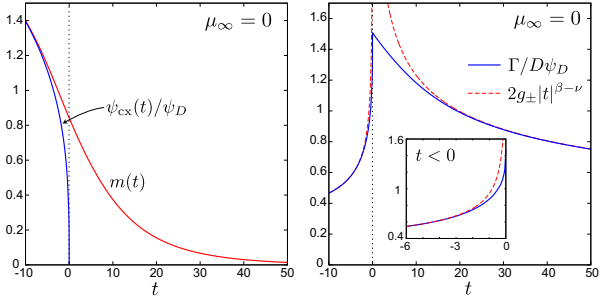}
\caption{\protect  
(Color online) 
Left: Normalized midpoint order 
parameter $m(t)=\psi_m(t)/\psi_D$ vs 
$t=\tau (D/\xi_0)^{1/\nu}$ 
for a film with  thickness $D$,  where 
$s=0$ for $\tau>0$ and 
$s= \psi_{\rm cx}/\psi_{\rm cx}$ 
for $\tau<0$ ($\mu_{\infty}=0$). The coexistence curve 
 $\psi_{cx}(t)/\psi_D$ vs $t$ 
  in Eq.(3.3)  is also shown in the region $t<0$.   
 Right: Normalized excess adsorption $\Gamma/D\psi_D$ vs 
$t$ for  $\mu_\infty=0$, whose maximum is $1.51$ at $t=0$. 
It  tends to  
$2g_{\pm}|t|^{\beta-\nu}$ for $t\gs 10$ 
and for $t\ls -1$ 
with $g_+=1.245$ and $g_-= 0.471$, respectively (see Eq.(3.45)). 
}
\end{center}
\end{figure}

\begin{figure}[htbp]
\begin{center}
\includegraphics[scale=0.69]{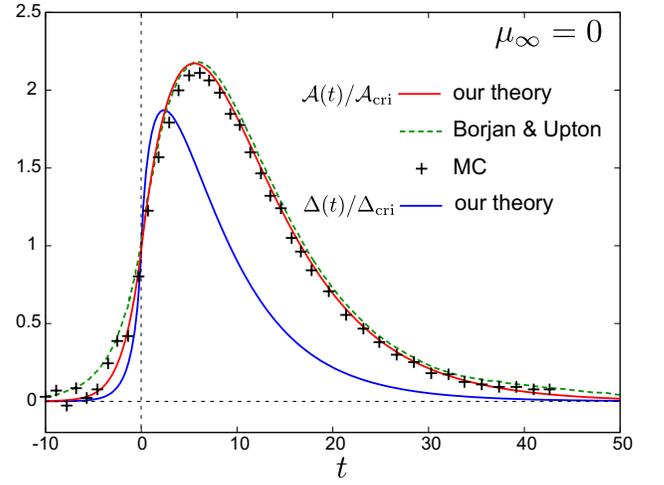}
\caption{\protect 
(Color online) 
${\cal{A}}(t)/{\cal{A}}_{\rm cri}$ 
vs $t=\tau (D/\xi_0)^{1/\nu}$ along the critical path $\mu_\infty=0$  
from our theory (red bold line) 
and from Borjan and 
Upton's theory \cite{Upton1} (blue broken line), 
which are in good agreement  with 
the Monte Carlo data ($+$) \cite{Monte}. 
Also plotted is  $\Delta(t)/\Delta_{\rm cri}$ 
vs $t$ from our theory on the critical path. 
}
\end{center}
\end{figure}

\begin{figure}[htbp]
\begin{center}
\includegraphics[scale=0.44]{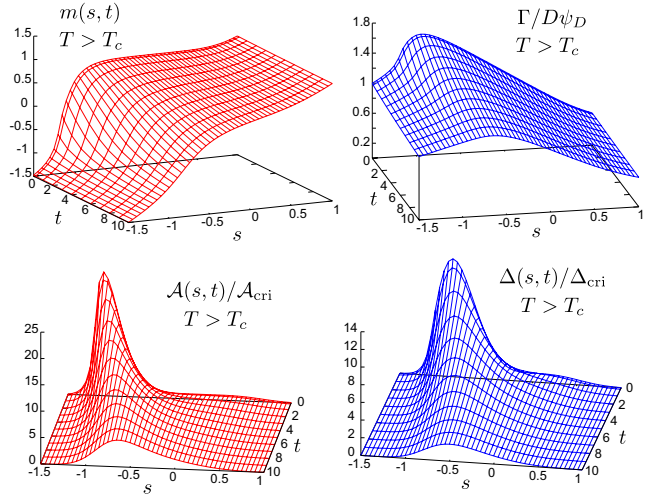}
\caption{\protect 
(Color online) 
Normalized midpoint order parameter 
$m(s,t)= \psi_m /\psi_D$ (top left), 
normalized excess adsorption $\Gamma/\psi_D D$ (top right), 
 ${\cal{A}}(s,t)/{\cal{A}}_{\rm cri}$ (bottom left), 
and   $\Delta(s,t)/\Delta_{\rm cri}$  (bottom right) for $t>0$ 
in the $s$-$t$ plane. 
The latter three quantities 
 are peaked for $s \sim -1$. The 
 amplitudes  ${\cal{A}}(s,t)$  and 
  $\Delta(s,t)$ are  very small 
for $t \gg 1$ and $|s|\gg 1$. 
}
\end{center}
\end{figure}

In our theory, 
Eqs.(2.36)-(2.39) 
and Eqs.(2.49)-(2.51) remain valid even  for $t\neq 0$. 
We now treat $m$, ${\cal A}$, $\Delta$, 
and $\Gamma^*$ as functions of $s$ and $t$, 
so $\p(\cdots) /\p s=(\p (\cdots)/\p s)_t$ and 
$\p(\cdots) /\p t=(\p (\cdots)/\p t)_s$. 
From Eq.(2.38) we obtain 
the generalized form of Eq.(2.48) as   
\be  
\frac{1}{R_\infty} \frac{\p\cal A}{\p s}=
\frac{\mu_m-\mu_\infty}{\psi_D 
k_BT_c}{\chi_\infty} \frac{\p m}{\p s}
- (m-s).
\en 
Here,  $R_\infty$ is a function of $s$ and $t$ 
defined by Eq.(2.52) (see Fig.16 for its 
overall behavior in the $s$-$t$ plane).  
Although redundant, we again write Eq.(2.51) as 
\be  
\frac{1}{R_\infty}  \frac{\p \Delta}{\p s}= \Gamma^*, 
\en 
which  follows from  the Gibbs adsorption relation. 
Here, $\Gamma^*$ is defined by Eq.(2.35), 
where the excess adsorption 
in the semi-infinite case $\Gamma_{{\rm sem}\pm}$ 
should be calculated for each given $\psi_\infty$ 
and $\tau$. 
In addition, differentiation of 
Eq.(2.49) with respect to $D$ yields the generalization 
 of Eq.(2.53): 
\be 
 {\cal A}= 
(d-1){ \Delta}-\frac{\beta s}{\nu} \frac{\p \Delta}{\p s}- 
 \frac{t}{\nu} \frac{\p \Delta}{\p t} .   
\en

\subsection{Results 
for $T>T_c$ }

We first give some analysis  along 
 the critical path $\mu_\infty=0$, where 
$\psi_\infty=0$ for $t>0$ 
and $\psi_\infty=\psi_{\rm cx}>0$ 
for $t<0$.  Numerical and experimental 
studies   on  the critical  adsorption 
and the Casimir amplitudes 
have  mostly been along this path in  the literature 
\cite{Ha,Monte,Krech,Lawreview,adDietrich,Upton,Upton1,Upton-ad}. 

Since  Eqs.(2.23)-(2.26) 
still hold,    $\psi(z)$ 
may be calculated  in the same manner as in 
Borjan and Upton's paper on 
  the critical  adsorption \cite{Upton-ad}. 
In Fig.6, we show  $\psi(z)/\psi_D$ 
for various  $t$ at $\psi_0/\psi_D=20$. 
For a film with large positive $t$, Eq.(2.26) 
gives $m=\psi_m/\psi_D $ in the form,  
\be 
m  \sim t^\beta\exp(-D/2\xi)\sim t^\beta\exp(-t^\nu/2).
\en  
For  $t \ll -1$,  Eq.(2.28) gives 
$m-s\sim s\exp(-D/2\xi) = s
\exp(-\xi_0|t|^\nu/2\xi_0')\ll 1$. 
In Fig.7, we plot 
$m(t)= \psi_m(t)/\psi_D$ 
and $\Gamma (t)/D\psi_D$ vs $t$. We can see that $m(t)$ 
decays as in Eq.(3.44) for $\tau\gg 1$ 
and tends to $\psi_{\rm cx}(t)/\psi_D$ 
for $\tau \ll -1$. As discussed around Eq.(2.35), 
 $\Gamma(t)$ tends to $2\Gamma_{{\rm sem}\pm}(t)$, 
where $\Gamma_{{\rm sem}\pm}$ 
is the excess adsorption in the semi-infinite case. 
In our theory  $\Gamma_{{\rm sem}\pm}$ behaves 
on the critical path as 
\be 
\Gamma_{{\rm sem}\pm}
=g_{\pm}' b_{\rm cx}\xi_0 |\tau|^{\beta-\nu}
= g_{\pm} 
|{t}|^{\beta-\nu} D\psi_D, 
\en 
where
$g_+= 0.66 g_+'= 1.245$ for $t>0$ and 
$g_-=0.66 g_-'= 0.471$ for $t<0$. 
We obtain the ratio 
$g_+/g_-= 2.64$, while it was estimated to be 2.28 
by  Fl\"{o}ter and  Dietrich \cite{adDietrich}. 
The right panel of Fig.7 shows that 
$\Gamma$ approaches the limit $2\Gamma_{{\rm sem}+}$ 
for $t \gs 10$ and $2\Gamma_{{\rm sem}-}$  
for $t \ls -1$.  For $t<0$, $\Gamma^*$ in Eq.(2.35) 
is very small. For example, it is   $ - 0.0675$ 
at $t=-1$ and  $ - 0.0122$ 
at $t=-2.2$ . 


 Figure 8 displays  the normalized  amplitudes 
${\cal A}(t)/{\cal A}_{\rm cri}$ 
and  ${\Delta}(t)/{\Delta}_{\rm cri}$ in our theory.  
We also  plot ${\cal A}(t)/{\cal A}_{\rm cri}$  
 from   the  Monte Carlo calculation  
 by Vasilyev {\it et al.} 
\cite{Monte} and from the local functional theory by 
Borjan and Upton \cite{Upton1}. 
Remarkably, the two  theoretical   curves  of 
${\cal A}(t)/{\cal A}_{\rm cri}$ 
excellently agrees with the Monte Carlo data 
in the region $t>0$. 
We should note that 
the free energy density in our theory 
and that in the linear parametric model \cite{Sc69}  
used by Borjan and Upton \cite{Upton1} 
are essentially the same 
on the critical path in the region $t>0$, 
as will be shown in Appendix B.  
In our theory, ${\cal A}/{\cal A}_{\rm cri}$  
and   ${\Delta}/{\Delta}_{\rm cri}$  
behave similarly.    
The maximum of  the former is  2.173 at $t=5.51$, 
while that of the latter   
is 1.872 at $t= 2.30$.

In the literature, however, 
there has been  no calculation 
of the Casimir amplitudes 
in the $s$-$t$ plane for $s \neq  0$ 
accounting for the renormalization effect. 
In  Fig.9, we display 
 $m(s,t)$,  $\Gamma (s,t)/\psi_D D$, ${\cal A}(s,t)$,  
and $\Delta (s,t)$  
for $t>0$ in the $s$-$t$ plane.  
We can see that ${\cal A}(s,t)$ 
and $\Delta (s,t)$ 
are both  peaked  at $s \sim -1$ 
and   behave similarly.

\subsection{Phase behavior of 
capillary condensation and enhancement of the 
Casimir amplitudes}

\begin{figure}[htbp]
\begin{center}
\includegraphics[scale=0.62]{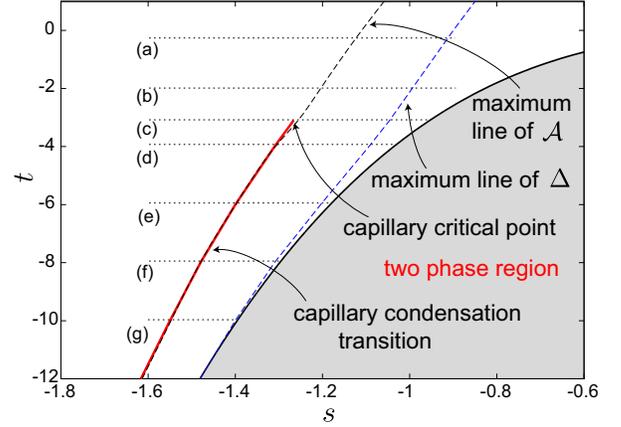}
\caption{\protect 
(Color online) 
Phase diagram   of a near-critical 
fluid  in a film for large adsorption in the $s$-$t$ plane, 
where $s=\psi_\infty/\psi_D(\propto 
\psi_\infty D^{\beta/\nu})$ and $t=\tau (D/\xi_0)^{1/\nu}$. 
Two-phase region is written in the right (in gray). 
There appears a 
first-order phase transition line (red) of 
capillary condensation with 
a critical point at $(s,t)= (-1.27, -3.14)$. 
Plotted also are  a line of maximum of $\Delta(s,t)$ 
with $(\p \Delta/\p s)_t=0$ and a line of maximum of 
${\cal A}(s,t)$ (broken lines). The former  approaches 
the coexistence curve for $t \ls -6$ and 
the latter is very close to the capillary condensation  line. 
On  paths (a)-(g), $m$, $\Gamma$, $\cal A$, and $\Delta$ 
are shown  in Figs.11 and 12. 
}
\end{center}
\end{figure}
\begin{figure}[htbp]
\begin{center}
\includegraphics[scale=0.6]{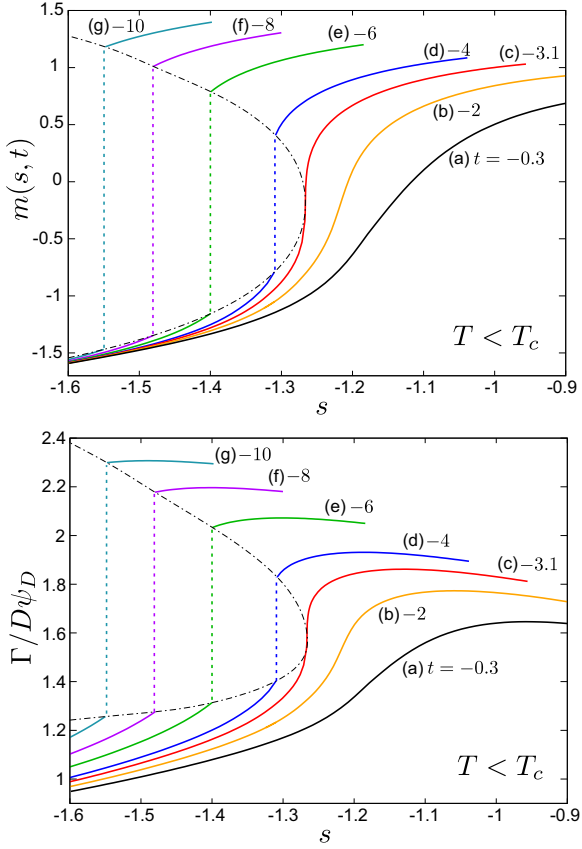}
\caption{\protect 
(Color online) 
 Normalized midpoint order parameter   
$m= \psi(D/2)/\psi_D$ (top) and normalized excess adsorption 
$\Gamma/\psi_D D$ (bottom) vs $s$   
in a  film with thickness $D$ 
for $t=-0.3, 
-2, -3.1, -4, -6, -8$, and $-10$ from the right. 
See  Eqs.(2.34) and (2.35) for  $\Gamma$.   
For $t<-3.1$,  these quantities are discontinuous 
across the  
capillary condensation line in Fig.10. 
}
\end{center}
\end{figure}

\begin{figure}[htbp]
\begin{center}
\includegraphics[scale=0.61]{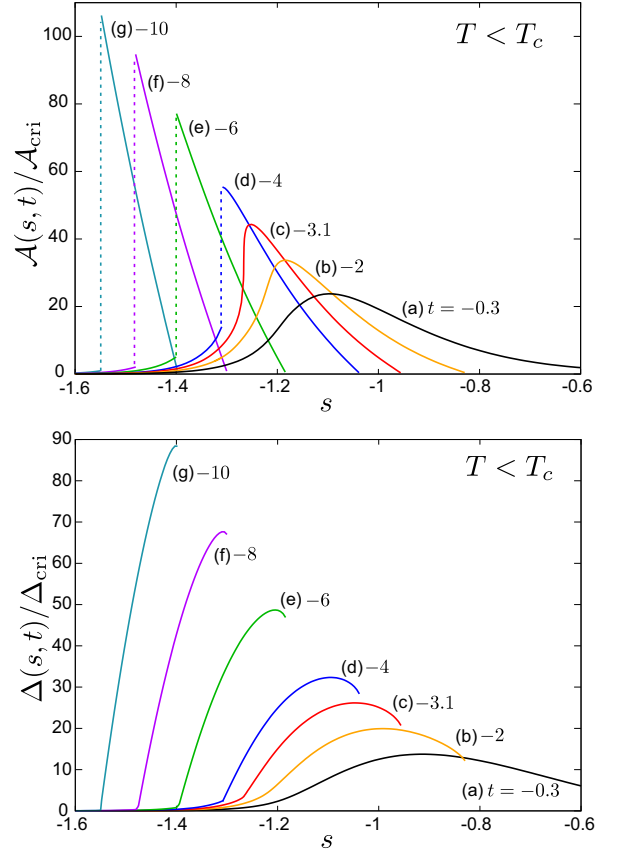}
\caption{\protect 
(Color online) 
Casimir amplitude ratios 
 ${\cal{A}}(s,t)/{\cal{A}}_{\rm cri}$ 
 (top) and  $\Delta(s,t)/\Delta_{\rm cri}$     
(bottom) vs $s$ 
for $t=-0.3, 
-2, -3.1, -4, -6, -8$, and $-10$ from the right. 
in the range $s< - \psi_{\rm cx}/\psi_D=-0.66 |t|^\beta$. 
Curves (c) corresponds to the capillary-condensation 
critical point. Those (d)-(g) exhibit a first-order phase transition, 
where ${\cal A}$ is discontinuous but $\Delta $ 
 is continuous.  
}
\end{center}
\end{figure}

\begin{figure}[htbp]
\begin{center}
\includegraphics[scale=0.44]{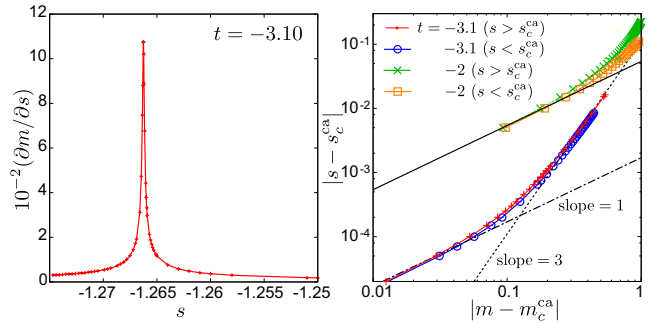}
\caption{\protect 
(Color online) 
Left: Susceptibility $(\p m/\p s)_t = 
(\p \psi_m/\p\psi_\infty)_\tau$ 
at $t=-3.1$ slightly above 
the capillary-condensation  critical point ($t_c^{\rm ca}=-3.14$).
Right: Relations between deviations $s-s_c^{\rm ca}$ 
and  $m-m_c^{\rm ca}$ 
near the capillary-condensation  critical point 
at $t=-3.1$ and away from it at $t=-2$ 
on a logarithmic scale. 
The former can be 
 fitted to the mean-field form (3.48). 
}
\end{center}
\end{figure}

We now discuss the 
phase behavior in the region of $t<0$ and $\psi_\infty<0$. 
In  Fig.10,  we  show 
 a first-order 
phase transition line 
 in the region  $t<0$ and $s<0$ 
outside the coexistence curve.   
As in Fig.11, 
the  discontinuities of the physical quantities  
across this line decrease 
with increasing $t$  
and vanish at a  critical point 
 $(\tau,\psi_\infty) =(\tau_c^{\rm ca},  
\psi_c^{\rm ca})$. 
In  agreement with the scaling 
theory \cite{Nakanishi-s},  we obtain  
\bea 
&&\tau_c^{\rm ca}= 
-3.14 (\xi_0/D)^{1/\nu}, \\ 
&& \psi_c^{\rm ca} = -1.27 \psi_D   \cong  -1.87 b_{\rm cx} 
(\xi_0/D)^{\beta/\nu}.
\ena  
We  use  the second line of Eq.(3.28) in Eq.(3.47). 
For $\tau<\tau_c^{\rm ca}$, let 
us write the transition point as 
 $\psi_\infty= \psi_{\rm ca}(\tau)$ 
or as $s=s_{\rm ca}(t)=  \psi_{\rm ca}(\tau)/\psi_D$.  
This  line divides the   condensed  phase 
with $m>0$ 
in the range   $s_{\rm ca}<s< -\psi_{\rm cx}/\psi_D$ 
and the  noncondensed phase 
 with $m<0$ in the range  $s<s_{\rm ca}$. 
In   Fig.11,  the isothermal 
curves of $m=\psi(D/2)/\psi_D$ 
and $\Gamma/D \psi_D$ are 
shown as functions of $s$ for $t<0$, 
where they are 
continuous for $t>t_c^{\rm ca}=-3.14$ 
and  discontinuous for $t<t_c^{\rm ca}$. 
It  is the capillary condensation  line 
for the gas-liquid transition 
\cite{Evansreview,Gelb} and is also 
the two-dimensional 
transition line for 
 Ising-like films 
\cite{JSP,Evans-Marconi,Nakanishi-s,Nakanishi-mean}.

In Fig.10, 
we also plot two dotted lines. 
On one line, 
 ${\cal A}(s,t)$ takes a maximum 
 as  a function of $s$ at fixed $t$. 
For $t<t_c^{\rm ca}$, it 
is slightly separated from the transition line 
 for $t>-4.0$ but coincides with 
 the transition line for $t<-4.0$. 
On the other  line close  to the  
bulk coexistence curve,  
 $\Delta(s,t)$ takes a maximum 
 as  a function of $s$ at fixed $t$ 
  and  we have  $(\p\Delta/\p s)_t=0$ 
  and  $\Gamma^*=0$ from Eq.(3.42).



In Fig.12, we plot the Casimir amplitudes 
${\cal A}(s,t)$ and 
$\Delta (s,t)$ as functions of $s$ 
for $t<0$.  For $t<t_c^{\rm ca}$, they grow 
very strongly in the condensed phase. 
As a marked feature,  ${\cal A}$ and 
$\Delta $ behave very differently 
for $t<0$, though they behave similarly for $t>0$ in Fig.9. 
These results are consistent with 
their derivatives with respect to $s$ 
in Eqs.(3.41) and (3.42).  
Previously, enhancement of $\cal A$  
was found  close to the transition line in  
the condensed phase  
by  Macio\l ek {\it et al.} \cite{Evans-Anna} 
in two-dimensional Ising films  
and by Schlesener {\it et al.} \cite{JSP} 
in the mean-field theory.

In Fig.11, the slope $ (\p m/\p s)_t$ 
or the susceptibility 
$dm/ds$  in Eq.(2.32) 
diverges  as $t$ is decreased to $t_c^{\rm ca}$. 
Thus, in  Fig.13,  we plot  $ (\p m/\p s)_t$ vs 
 $s$ for $t=-3.1$ in the left panel  and the curve of 
$m$ vs $s$ at $t=-3.1$ and -2 in the right panel. 
The curve of $t=-3.1$ can 
well  fitted to   the following mean-field form,     
 \be 
s-s_c^{\rm ca} =A_{\rm ca}(t-t_c^{\rm ca})( m-m_c^{\rm ca})  
 + B_{\rm ca}(m-m_c^{\rm ca})^3 ,
\en   
 where $(t_c^{\rm ca}, s_c^{\rm ca},m_c^{\rm ca})= (-3.14, -1.27, -0.172)$,  
 $A_{\rm ca} =0.0425$,  and $B_{\rm ca}=0.11$. 
This   mean-field  behavior near the capillary-condensation  
critical point  arises because 
 the long wavelength fluctuations of 
$\psi$ inhomogeneous 
in the lateral plane have been neglected.  
The curve of $t=-2$  
 is not well fitted to 
Eq.(3.48) for  $|m-m_c^{\rm ca}|\gs 0.3$.

\subsection{Determination of the  
capillary condensation line}

In  Fig.14, we show the profile  
 $\psi(z)$  at  $t= -4$ 
for four values of $s$ given by (A) $-1.10$, 
(B) -1.29, (C) $-1.33$,  and (D) -1.50, where 
 $\psi_0=20 \psi_D$. 
 The two lines  $\psi =\pm \psi_{\rm  cx}$ are also shown, 
 between which $|\psi|< \psi_{\rm  cx}$ 
and the free energy density is given by 
the mean-field form (3.31). 
Here, between (B) and (C), 
there is a first-order phase transition 
 at $s=-1.31$,  where  
the normalized midpoint value $m$ changes discontinuously 
between  $0.407$ and $-0.793$.

\begin{figure}[htbp]
\begin{center}
\includegraphics[scale=0.6]{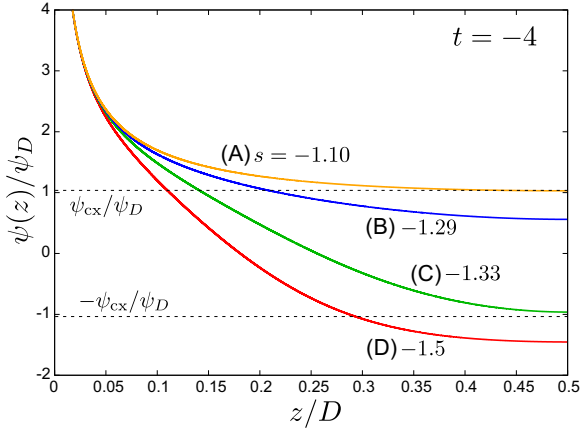}
\caption{\protect 
(Color online) 
 Normalized order parameter   
$\psi(z)/\psi_D$ vs $z/D$   
in a  film  in the half region  $0<z<D/2$ at  $t=-4$, 
with $\psi_0=20 \psi_D$ in the symmetric boundary 
conditions. 
From above,  (A) $s=-1.10$ (at maximum of $\Delta$), 
(B) -1.29 (near maximum of $\cal A$), 
(C) $-1.33$,  and (D) $-1.50$.  A 
first-order phase transition occurs 
between (B) and (C).  
 Lines of $\psi_\infty =\pm \psi_{\rm  cx}$ 
are written, where  $\psi_{\rm cx}/\psi_D=1.04$.}
\end{center}
\end{figure}

\begin{figure}[htbp]
\begin{center}
\includegraphics[scale=0.4]{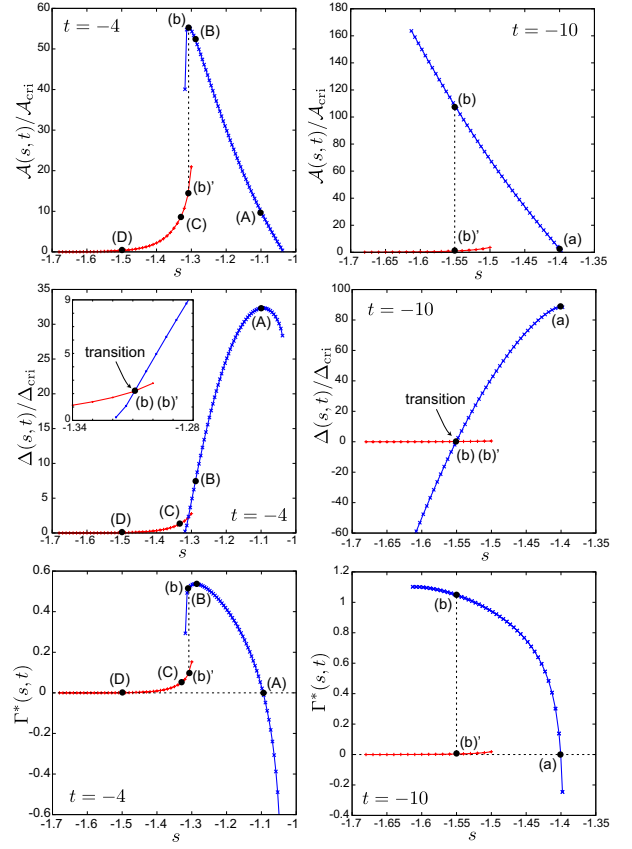}
\caption{\protect 
(Color online) 
${\cal{A}}(s,t)/{\cal{A}}_{\rm cri}$ (top), 
  $\Delta(s,t)/\Delta_{\rm cri}$ (middle), 
 and $\Gamma^*(s,t)$ (bottom) 
 for  $t=-4$ (left) 
 and $t=-10$ (right). 
 For $t=-4$, points (A), (B), (C), and (D) correspond to the 
curves in Fig.14. 
There are equilibrium 
 and metastable branches near the   transition. 
 Amplitude $\Delta$  is 
 maximized at (A) (left) and at (a) (right),  
where $\Gamma^*=0$ from Eq.(3.42). The transition 
 occurs between 
(b) and (b'), where $\Delta$ is continuous.   
}
\end{center}
\end{figure}

\begin{figure}[htbp]
\begin{center}
\includegraphics[scale=0.74]{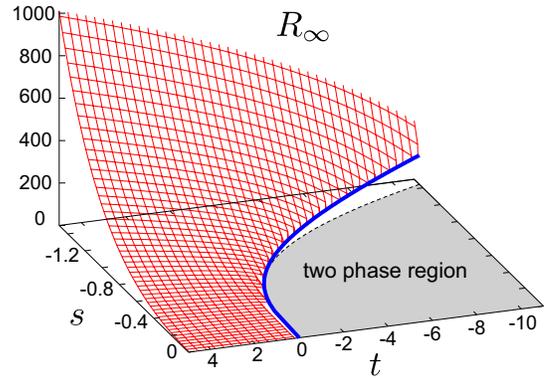}
\caption{\protect 
(Color online) 
Scaled inverse susceptibility for a film 
 $R_\infty$ in Eqs.(2.52) and (3.42) in the $s$-$t$ plane, 
 which is of order $40|s|^{\delta-1}$ for $|s| \gs 1$.  
}
\end{center}
\end{figure}


In  Fig.15,  ${\cal{A}}(s)/{\cal{A}}_{\rm cri}$, 
  $\Delta(s)/\Delta_{\rm cri}$,  
and $\Gamma^*$ 
 are displayed  as functions of  $s$ 
 at  $t=-4$ (left) and -10 (right), where 
 $\psi(z)$  for the points 
   (A), (B), (C), and (D)  at  $t=-4$ 
 can be seen   in Fig.14.  
In numerical analysis, 
we obtained two branches of the profiles 
giving rise to  hysteretic behavior  
at the transition. 
From Eq.(2.49) $\Delta$ should be 
maximized  in equilibrium. 
Thus the equilibrium (metastable) 
branch should be the one with   a larger 
(smaller) $\Delta$. A first-order phase transition occurs 
at a point where  the two curves  of $\Delta (t)$ cross. 
We confirm 
 the following. (i) In  accord with Eq.(3.42), 
 the maximum point of $\Delta$ coincides with 
the vanishing point of $\Gamma^*$, as can be known from 
comparison of  the middle and bottom panels. 
(ii) As $s \to -\psi_{\rm cx}/\psi_D$, 
 ${\cal A}$ becomes very small on 
 the top plates, since it 
 approaches the value on 
 the critical pass $\mu_\infty=0$ with $t<0$ 
 displayed in Fig. 8 (see the sentence below Eq.(3.44)). 
(iii) As $s$ is decreased from 
  $-\psi_{\rm cx}/\psi_D$ to 
  the transition  value 
  $s_{\rm ca}$ in 
  the condensed phase,  
${\cal A}$ grows with a steep negative  
slope.  To understand this behavior, 
we compare the first and second terms 
in the right hand side of Eq.(3.41). 
We notice that the ratio of the 
first term to the second   
 is very small  in this $s$ range.  
 It  is between 
$[0.02,0.05]$ for $t=-4$ and 
between  $[0.001,0.005]$ for $t=-10$. In 
the condensed phase 
in the range 
$s_{\rm ca}<s<-\psi_{\rm cx}/\psi_D$,  
 we thus find 
\be  
 \frac{\p\cal A}{\p s}\cong - R_\infty(m-s).
 \en 
While   $m-s$ remains  of order 
$\psi_{\rm cx}/\psi_D \sim |t|^\beta$,  
the  coefficient  $R_\infty$ is very large 
outside the coexistence curve.  
Figure 16 displays  the 
overall behavior of 
 $R_\infty$ in the $s$-$t$ plane, 
 where  $R_\infty \cong 6.6 |t|^\gamma$ 
on the coexistence curve and $R_\infty \cong 
A_c \delta(1+\delta) |s|^{\delta-1}\sim  
40 |s|^{\delta-1}$ for $|s| > 1$.

We also  comment on the validity of the general 
relation (3.43) in our numerical analysis. 
For example,  at $t= -4$ in Fig.15, 
let us consider the two points  (A) $s=-1.1$ 
and (B)  $s=-1.29$.   
 At point (A),  we have 
${\cal A}=5.54$, $\Delta= 9.02$, 
$\p \Delta/\p s= 3.13$, 
and $\p \Delta/\p t= -2.21$. 
The three terms in the right hand side of Eq.(3.43)  are 
thus $18.04$, $1.78$, and  $-14.03$ in this order 
and indeed their sum gives $\cal A$.   
At point (B), we have 
${\cal A}=23.98$, $\Delta= 1.98$, 
 $\p \Delta/\p s= 72.4$, and $\p \Delta/\p t= -3.85$, 
 so the three terms in the right hand side of Eq.(3.43) are  
 $3.96$, $-24.4$, and $48.2$ in this order, 
 whose sum indeed yields  $\cal A$.  

\section{Summary and remarks}
\setcounter{equation}{0}

We  have 
calculated the order parameter 
profiles and the Casimir amplitudes 
for a film of near-critical fluids. 
Our results are also applicable to one-component fluids 
near the gas-liquid critical point where  
 the walls favor  either of gas or liquid. 
We summarize our main results.\\ 
(i) In Sec.II, we have used the singular 
free energy by 
Fisher and Au Yang  at   $T=T_c$. 
Using this model, we have    defined  
the two Casimir amplitudes,  
$\cal A$ for  the force 
density and  $\Delta$ for  the grand potential, 
 as functions of 
the scaled order parameter 
 of the reservoir $s=\psi_\infty/\psi_D$ in Eq.(3.31),  
They are sharply peaked at $s \sim -1$ 
and the peak heights are much larger than 
their critical-point values ${\cal A}_{\rm cri}$ 
and ${\Delta}_{\rm cri}$   as in Fig.4.  
These singular behaviors 
have been  analyzed analytically. 
This off-critical 
behavior   may also be interpreted as 
pretransitional enhancement,  
 because the region of $s \sim -1$ 
and $t=0$ is close to 
the capillary-condensation critical point.
\\ 
(ii) In Sec.III,  
we have constructed a  free energy 
with the gradient contribution for $T\neq T_c$ 
including the renormalization effects, with  which we may readily calculate 
the physical quantities. 
The Casimir amplitudes $\cal A$ and $\Delta$ 
are much amplified for $s \sim -1$ 
as shown in Fig.9 for $T>T_c$ and 
in Fig.12 for $T<T_c$. Their maxima are 
larger than their critical-point values by 10-100 times. 
We have then found a first-order phase transition line 
of capillary condensation for negative $t$ 
slightly outside the bulk coexistence curve, 
where the profile of $\psi$ and  $\cal A$ are 
discontinuous but $\Delta$ is continuous. 
This line ends at a critical point given by 
$(s,t)= (-1.27,-3.14)$. 
The amplitude $\cal A$ exhibits a maximum 
close to this line, while the amplitude $\Delta$ 
close to  the bulk coexistence curve. 
\\

We make some  remarks.\\
1) Even at the  mean-field  level,  it follows 
 the power-law form of the 
interaction free energy $\Delta F$ for $\xi > D> \ell_0$:  
\be 
\Delta F= -k_BT_c  D^{-4} \Delta_0.  
\en 
Here,  setting  $\delta=3$ 
in Eq.(2.3) and  $\eta=0$ in Eq.(2.4), we 
obtain   $\Delta_0
= 4I_0^4 C_0^2/3B_0k_BT_c$ 
at the criticality under strong adsorption, 
 where  $I_0$ is given by Eq.(2.45). 
The adsorption-induced interaction is already present 
in the mean-field theory\cite{Okamoto,Tsori} 
and   its form becomes universal 
near the criticality  \cite{Casimir}.  
  Enhancement of 
$\Delta F$  
  near the capillary condensation 
transition  is rather obvious in view of the fact 
that it occurs even in the mean-field theory \cite{JSP}. 
Note that $\Delta F \propto h_1^2$ 
for  weak adsorption $(\ell_0 >D$) away from the criticality, 
where $h_1$ is the surface 
field \cite{Okamoto}. 
\\  
2)
We have neglected the fluctuations varying in the lateral plane 
with wavelengths longer than the three-dimensional $\xi$. 
Thus the   capillary condensation transition has been treated 
at  the  mean-field level, leading to   
 Fig.13. In idealized  conditions,  
  there should be  
composition-dependent   crossovers  from  the Ising 
behavior in three dimensions to that in two dimensions.\\ 
3) Nucleation  and spinodal decomposition 
should take place  between plates and in porous media  
if $T$ is changed  across the capillary 
condensation line  outside the solvent 
coexistence curve \cite{Evansreview,Gelb}.  
\\ 
4) For neutral colloids, the attractive 
interaction arises  from overlap of 
composition deviations  near  the   
colloid  surfaces.  It is intensified 
if  the component favored by the  surfaces  
is  poor  in  the reservoir. 
A  bridging transition further takes place 
 at lower $\tau$ between  strongly 
 adsorbed  or wetting  layers 
 of colloid particles 
 \cite{Evans-Hop,Nature2008,Okamoto}. 
  \\
5) Our local functional theory  
has been used when $\psi$ varies  
over a wide range in strong adsorption. 
It can  be used in various  situations. 
For example, dynamics of colloid 
particles in near-critical fluids 
can be studied including the hydrodynamic flow.  
So far,  phase separation in near-critical fluids 
has been investigated 
in the scheme of  the $\psi^4$ theory 
 with constant renormalized coefficients 
\cite{Onukibook}.  However, the distance to 
the bulk criticality (the parameter $w$ in Sec.III) 
can be  inhomogeneous around 
preferential walls or around heated or cooled walls.\\ 
6) We should further investigate 
the ion effects in confined 
multi-component fluids.  
In such situations, 
the surface  ionization can  depend on  the ambient  
ion densities and  composition  
\cite{Current,Okamoto}. A prewetting transition then 
appears even away from the 
solvent coexistence curve, where  the degree of 
ionization 
is also discontinuous. 
We are interested in how  the 
ionization fluctuations   
affect the ion-induced 
capillary condensation transition 
\cite{Tsori}.  
\\

\begin{acknowledgments}
This work was supported by Grant-in-Aid 
for Scientific Research  from the Ministry of Education, 
Culture,  Sports, Science and Technology of Japan. 
One of the authors (A.O.) would like to thank 
Daniel  Bonn  and   Hazime  Tanaka for 
informative correspondence. 
\end{acknowledgments}

\vspace{2mm}
\noindent{\bf Appendix A: 
Mechanical equilibrium}\\
\setcounter{equation}{0}
\renewcommand{\theequation}{A\arabic{equation}}
 
In one-dimensional situations, 
the $zz$ component of the stress tensor 
due to the composition deviation is \cite{Onukibook} 
\be 
\Pi_{zz}= \psi \mu- f 
+C (|\psi'|^2/2-\psi\psi'') -  C' 
\psi|\psi'|^2/2, 
\en 
where $C'= dC/d\phi$. 
The  mechanical equilibrium condition is  $d \Pi_{zz}/dz=0$, 
 so $\Pi_{zz}$  is a constant 
independent of $z$. 
We further use Eq.(2.11) 
to eliminate the term proportional to $\psi''$ 
to obtain 
\be 
\Pi_{zz}= \psi \mu_\infty - f 
+C |\psi'|^2/2.
\en  
Thus $\Pi_{zz}= \psi_m\mu_\infty -f_m$ at $z=D/2$. 
The osmotic pressure is  given by 
\be 
\Pi= \Pi_{zz}- (\psi_\infty \mu_\infty-f_\infty),
\en  
so we find   $\Pi = -\p \Omega/\p D$ in  Eq.(2.39).

\vspace{2mm}
\noindent{\bf Appendix B: 
Relationship  to the Schofield, Lister, and Ho 
linear parametric model }\\
\setcounter{equation}{0}
\renewcommand{\theequation}{B\arabic{equation}}

The linear parametric model 
\cite{Sc69} 
provides  the equation of state  and   thermodynamic 
quantities   of Ising   systems 
 in  compact forms \cite{Hohenberg,Wallace2} 
 for detailed  discussions  on this model. 
 It uses  two parametric  
variables, $\hat{r}$ 
and $\theta$, with   $\hat{r} \ge 0$ and $|\theta|
 \le 1$;  $\hat{r}$ represents a  distance  
 to the critical point 
and $\theta$ an  angle around it 
in the $\psi$-$\tau$ plane. 
Here $\hat r$ should not be confused with $r$ in Eq.(3.5). 
In this model, 
homogeneous equilibrium states are supposed.
The reduced temperature 
$\tau$, the magnetic  field  
$h$, and the average order parameter $\psi$  
are expressed in terms of  $\hat r$ and $\theta$ as   
\bea
\tau &=& {\hat r} (1-b_c^2\theta^2) ,\\
h &=&  a_{0}  \hat{r}^{\beta\delta} \theta (1-\theta^2),\\
\psi&=&  c_0 {\hat r}^\beta \theta. 
\ena
Here $a_0$ and $c_0$ are positive nonuniversal constants,  
while  $b_c$ is a universal number.
The case $\theta=0$ corresponds to $\tau>0$ and $h=0$, 
$\theta=\pm 1/b_c$ to $\tau=0$,  and $\theta=\pm 1$ to the 
coexistence curve ($h=0$ and $\tau <0$).
We may calculate various thermodynamic 
quantities from these relations in agreement with 
 the asymptotic critical behavior. 
Though  $b_c$ is arbitrary within the model, 
 $b_c^2$ was set equal to 
\be 
b_c^2= (\delta-3)/[(\delta-1)(1-2\beta)] \cong 1.36.
\en 
This choice  yields simple 
expressions for  the 
critical amplitude ratios  in close 
agreement  with experiments. 
The linear parametric model 
 in  Eqs.(B1)-(B4) is exact up to order 
 $\epsilon^2$ \cite{Wallace2}. The  
 two-scale-factor universality 
 \cite{Stauffer,HAHS} furthermore indicates that 
the combination $(a_0c_0)\xi_{0}^d$ 
of the coefficients in Eqs.(B2) and (B3) 
should be  a universal number, where $\xi_0$ 
is the microscopic length in 
the correlation length 
$\xi=\xi_0\tau^{-\nu}$ for $\tau>0$ and $h=0$.

Our model in Eqs.(3.4)-(3.9) closely resemble 
the liner parametric model as regards the 
thermodynamics of homogeneous states. 
Note that 
$h$ in Eq.(B2) corresponds to $\mu/k_BT$ in Eq.(3.20) 
and $\hat r$ to $w$ in Eq.(3.9). 
For our model, 
we may introduce the angle variable $\theta$ by 
\be 
\theta= {\rm sign}(\psi) b_O^{-1} \sqrt{1-S},
\en 
where $S$ is defined in Eq.(3.17). 
We set  $\theta=\pm 1$ on the coexistence curve so that 
\be 
b_O^2={1+1/\sigma}\cong 1.58,
\en 
where $\sigma$ is given by Eq.(3.22). 
Then $\tau$, $\mu$, and 
$\psi$ are  expressed in terms of $w$ and  $\theta$ as 
\bea 
&&\tau= w (1-b_O^2\theta^2),\\
&&\frac{\mu}{k_BT}= C_3 w^{\beta\delta}  \theta 
(1-\theta^2)\frac{1- A_1 \theta^2}{1-A_2\theta^2},\\ 
&& \psi= (b_O /\sqrt{C_2})  w^\beta\theta, 
\ena 
where the coefficients $A_1$, $A_2$, and $C_3$ are expressed as 
\bea 
A_1&=& 5\alpha b_O^2/[5\alpha+(2-\alpha)\sigma]\cong 0.23, \\
A_2&= &(1-2\beta) b_O^2\cong 0.55, \\ 
C_3&=& b_O^2 [5\alpha+(2-\alpha)\sigma]C_1/[6\sqrt{C_2}\xi_0^2].
\ena 
As  differences 
between our model and  the 
linear parametric model, 
 $b_O^2$ in Eq.(B6) is  larger 
than $b_c^2$ in Eq.(B3) by $16\%$ 
and there appears the extra  factor, 
\be 
Z(\theta) =  
({1- A_1 \theta^2})/({1-A_2\theta^2}), 
\en  
in the right hand side of Eq.(B8). 
Our model considerably deviates from the 
parametric model 
close to the coexistence curve mainly because of  $Z(1)=1.73$.  
The correlation length $\xi$ 
and the susceptibility $\chi^{-1}$ on the coexistence curve 
are thus underestimated in our theory as in Eqs.(3.26) and (3.27). 
Also in our model the combination   
$C_3 (b_0/\sqrt{C_2})\xi_0^d$ 
for the coefficients in Eqs.(B8) and (B9) 
is a universal number from Eq.(3.11) in accord with the  
 two-scale-factor universality.


\begin{references}
\bibitem{chemical} 
Binary mixtures are characterized by the temperature  
$T$ and the chemical potential difference 
$\mu= \mu_1-\mu_2$ betwen the two components  
 at a constant pressure $p$. 
In the text,  $\mu_\infty=\mu- \mu_c$ in the resevoir, where 
$\mu_c$ is the critical value.
\bibitem{Casimir} 
The long-range interaction ($\propto D^{-d+1}$) 
in near-critical binary mixtures treated in this paper 
arises from 
the slow and universal composition decay  
in the fluid induced by the  wall perturbation. 
From our viewpoint, it is  somewhat  misleading to 
call it  the Casimir interaction, though we use 
"Casimir amplitudes".  
The Casimir interaction itself was originally 
found to be induced 
by the ground-state fluctuations  
of the electromagnetic field between two mirrors.
\bibitem{constant} 
The singular free energy 
density is of the form  
$f_{\rm s} = -k_BT_c\xi_0^{-d}
 A_{+}\tau^{2-\alpha}$ 
 for $\tau>0$ and $\psi=0$, where 
 $A_{+}$ is  a universal number of order 0.1 
  \cite{Onukibook,HAHS}. 
 Twice differentiating $f_{\rm s}$ with respect to $T$, 
 we obtain the well-known specific heat 
 $C= k_B\xi_0^{-d}(2-\alpha)(1-\alpha) 
 A_{+}\tau^{-\alpha}$.
\bibitem{sus} 
The susceptibility $\chi$
is given by  $k_BT_c\psi/\mu \chi= 
1+ (1-\alpha)\varphi+\beta(1-2\beta)\varphi^2 - 
[4(1-\alpha)+ 10\alpha S]S\varphi/[
 2-\alpha +4(1-\alpha)S+5\alpha S^2]$, 
 where $\varphi=2(1-S)/[2\beta+ (1-2\beta)S]$.
\end{references}
\end{document}